\documentclass[10pt,leqno]{amsart}
\usepackage{lineno}
\usepackage[foot]{amsaddr}
\usepackage{amsmath}
\usepackage{graphicx,psfrag,epsf}
\usepackage{enumerate}
\usepackage{natbib}
\usepackage{url} % 

\usepackage{mathtools, empheq}
\usepackage{amssymb, setspace}
\usepackage{bm, bbm}
\usepackage{graphicx, tabularx, array}
\usepackage[svgnames,table]{xcolor}
\usepackage{enumitem}
\usepackage{multirow}
\usepackage{hyperref}
\usepackage{makecell}
\usepackage{booktabs}
\usepackage{amsmath}
\usepackage{array}

\usepackage{longtable}

\usepackage{colortbl} % 

\newcolumntype{C}[1]{>{\centering\arraybackslash}m{#1}}

\usepackage{mathtools}

\DeclarePairedDelimiter\floor{\lfloor}{\rfloor}

\newcommand{\E}{\mathbb{E}}

\newcommand{\Z}{\mathcal{Z}}
\newcommand{\F}{\mathcal{F}}
\newcommand{\G}{\mathcal{G}}

\renewcommand{\P}{\mathbb{P}}
\newcommand{\V}{\text{var}}
\newcommand{\sgn}{\text{sgn}}
\newcommand{\1}{\mathbbm{1}}

\renewcommand{\c}{\text{con}}
\newcommand{\p}{\text{plan}}
\newcommand{\cv}{\text{coverage}}
\renewcommand{\a}{\text{analysis}}

\theoremstyle{plain}
\newtheorem{theorem}{Theorem}
\newtheorem{method}{Method}

\newtheorem{proposition}{Proposition}
\newtheorem{corollary}{Corollary}

\usepackage{newtxtext}
\usepackage[subscriptcorrection]{newtxmath}

\graphicspath{{./art/}}

\usepackage[plain,noend]{algorithm2e}

\makeatletter
\renewcommand{\algocf@captiontext}[2]{#1\algocf@typo. \AlCapFnt{}#2} % text of caption
\def\@algocf@capt@plain{top}
\renewcommand{\algocf@makecaption}[2]{%
  \addtolength{\hsize}{\algomargin}%
  \sbox\@tempboxa{\algocf@captiontext{#1}{#2}}%
  \ifdim\wd\@tempboxa >\hsize%     % if caption is longer than a line
    \hskip .5\algomargin%
    \parbox[t]{\hsize}{\algocf@captiontext{#1}{#2}}% then caption is not centered
  \else%
    \global\@minipagefalse%
    \hbox to\hsize{\box\@tempboxa}% else caption is centered
  \fi%
  \addtolength{\hsize}{-\algomargin}%
}
\makeatother

\begin{document}

\def\spacingset#1{\renewcommand{\baselinestretch}%
{#1}\small\normalsize} \spacingset{1}

\title[Hypothesis screening with split samples in matched observational studies]{Planning for gold: Hypothesis screening with split samples for valid powerful testing in matched observational studies}

\author{William Bekerman$^{*1}$} \email{bekerman@wharton.upenn.edu}
\author{Abhinandan Dalal$^{*1}$}
\email{abdalal@wharton.upenn.edu}
\author{Carlo del Ninno$^{2}$}
\author{Dylan S. Small$^{1}$}
\email{dsmall@wharton.upenn.edu}

\dedicatory{$^{1}$Department of Statistics and Data Science, University of Pennsylvania, Philadelphia, PA, USA\\$^{2}$Retired, Formerly of the World Bank, District of Columbia, USA}

\thanks{$^*$Denotes equal contribution}

\begin{abstract}
Observational studies are valuable tools for inferring causal effects in the absence of controlled experiments. However, these studies may be biased due to the presence of some relevant, unmeasured set of covariates. One approach to mitigate this concern is to identify hypotheses likely to be more resilient to hidden biases by splitting the data into a planning sample for designing the study and an analysis sample for making inferences. We devise a powerful and flexible method for selecting hypotheses in the planning sample when an unknown number of outcomes are affected by the treatment, {allowing researchers to gain the benefits of exploratory analysis and still conduct powerful inference under concerns of unmeasured confounding}. We investigate the theoretical properties of our method and conduct extensive simulations that demonstrate pronounced benefits, especially at higher levels of allowance for unmeasured confounding. Finally, we demonstrate our method in an observational study of the multi-dimensional impacts of a devastating flood in Bangladesh.
\end{abstract}


\maketitle

\spacingset{1.5}

% \tableofcontents

\section{Design of Observational Studies and Data Splitting}

The design of an observational study may encompass various considerations, including what protocol to use, the level of unconfoundedness at which to control, choice of test statistics, and outcomes for which to test. \cite{rosenbaum2010design} argues that a methodical design strengthens the power and transparency of observational analyses and that a glimpse of the data is often advantageous. Randomly splitting the sample into a smaller planning sample for choosing the design and a larger analysis sample for making inferences can help maintain an honest and transparent inference process, while not sacrificing validity (\cite{heller2009split, rosenbaum2010design, small2024protocols}). 
In practice, a researcher would aim to be deliberate in how she allocates the power of her data analysis, preferring to select hypotheses which would have a reasonable chance of being significant in the analysis sample, even after accounting for potential confounding biases. The goal of this paper is to develop a powerful method for selecting outcomes to test on the analysis sample from the planning sample data. As a motivating example, we consider the wide-ranging effects of the 1998 floods in Bangladesh, which endangered the health and lives of millions through food shortages, diminished purchasing power for essential goods, and the potential spread of water-borne illnesses (\cite{del20011998}). We employ split-sampling to establish hypotheses likely to be less sensitive to unmeasured biases, helping researchers to identify the specific impacts of treatment; in this case, evaluating whether these floods reduced food availability, decreased access to sanitary water, or worsened illness intensity, to highlight a few possibilities, among residents in Bangladesh following these disasters. %

Seeking to improve design under concerns of potential unmeasured confounding, \cite{heller2009split} utilize sample splitting to compare design sensitivities (\cite{rosenbaum2004design}). They demonstrate that sample splitting may reduce power, but leaves design sensitivity unchanged. The most salient rebuttal against the use of data splitting has been the loss of power resulting from the decrease in sample size. \cite{cox1975note} utilized sample splitting in randomized trials to choose several hypotheses. Similar approaches have been employed by \cite{wasserman2006weighted}, weighting hypotheses based on estimated signal strengths, and \cite{rubin2006method}, estimating optimal thresholds in a multiple testing problem. Unfortunately, the gain in power using such clever techniques has generally failed to compensate for the reduced sample size in randomized experiments. Yet \cite{heller2009split} show that sample splitting can be powerful in the context of observational studies. Although their work constitutes an illuminating proof of concept, their approach of choosing the smallest $p$-value is largely heuristic-based and is limited to either one specific outcome or one general outcome of interest which can manifest itself through various avenues. In his defense of the use of protocols for observational studies, \cite{small2024protocols} (Section 6.1) raises the choice of how many and which hypotheses to select in a planning sample as an open problem.

In this article, we provide an adaptive procedure to select hypotheses to test in an observational study with many outcomes. Our method is based on the principle of choosing outcomes that are more robust to potential hidden biases and relies on the sensitivity value, coined first by \cite{zhao2018sensitivity} but frequently reported in observational studies since \cite{cornfield1959smoking}. Defined as the smallest level of unmeasured confounding $\tilde\Gamma^l$ required in sample to invalidate the significance of a causal effect, the sensitivity value is a sample-based random variable that captures the trade-off between finite sample efficiency and bias insensitivity more succinctly than the design sensitivity. In more detail discussed in the subsequent sections, our strategy has the advantage of dealing with the behavior of a finite sample, as it is based on the random variable (sensitivity value) rather than its asymptotic limit parameter (design sensitivity). In particular, our method computes the sensitivity value of each outcome on the planning sample and estimates its corresponding variability to construct predictive intervals on the analysis sample. Our approach demonstrates competitive power compared to the full sample Bonferroni correction in observational studies, while allowing for the full benefits of data splitting to be reaped. Simulation results suggest that benefits are more pronounced with increasing wariness about unmeasured confounding; that is, increasing $\Gamma$-levels at which to control.

\section{Hidden Bias in Matched Observational Studies} \label{notation}

\subsection{Treatment assignments and treatment effects}
Consider an observational study with $I$ matched sets, $i=1,\ldots, I$, where the $i$th set contains $n_i\ge 2$ subjects, $j=1,\ldots, n_i$, one of which is treated with $Z_{ij} = 1$, while the others are controls with $Z_{ij} = 0$, such that $\sum_{j=1}^{n_i} Z_{ij} = 1$ and $N = \sum_{i=1}^{I} n_i$ subjects in total. Sets are matched for observed covariates $x_{ij} = x_{ij'}$ for all $i, j, j'$, but a researcher may be concerned about an unobserved covariate possibly leading to $u_{ij}\ne u_{ij'}$ for some $i, j, j'$. There are $L$ outcomes in which the researcher may be interested. Let $r^l_{T_{ij}}$ and $r^l_{C_{ij}}$ denote the potential outcomes for the $l$th outcome for the $j$th subject in the $i$th matched set when assigned to treatment and control, respectively, for $l=1,\ldots, L$. The researcher gets to observe $R^l_{ij} = Z_{ij}r^l_{T_{ij}} + (1-Z_{ij})r^l_{C_{ij}}$; hence, the individual treatment effect $r^l_{T_{ij}} - r^l_{C_{ij}}$ is not calculable \cite{neyman1923application, rubin1974estimating}.  Take $\mathcal{F} = \{(x_{ij}, u_{ij}, r^l_{T_{ij}}, r^l_{C_{ij}}) \text{ 
 } | \text{  } i=1,\ldots, I; \text{  } j=1,\ldots, n_i; \text{  }  l=1,\ldots, L\}$ and $\mathcal{Z}$ be the event $\sum_{j=1}^{n_i} Z_{ij} = 1$. Let $\bm R^l = (R^l_{11},  \ldots, R^l_{In_I})^T$, $\bm R = (\bm R^1,\ldots, \bm R^L)^T$, and $\bm Z = (Z_{11}, \ldots, Z_{In_I})^T$. We use $\Omega$ to denote the support of $\bm Z$. Throughout the paper, $\1(A)$ refers to the indicator function, which takes value one if $A$ occurs and zero otherwise. 

In a randomized experiment, a subject within each matched set $i$ is randomly selected to receive treatment with probability $1/{n_i}$. Treatment assignment is independent across distinct matched sets in this setting, ensuring that $\mathbb{P}(z) = {1}/{|\Omega|}$ for all $z\in\Omega$, where $|\cdot |$ refers to the size of a finite set (\cite{fisher:1935}). The sharp null posits $H_0^l: r^l_{T_{ij}} = r^l_{C_{ij}}$ for all $i,j$. An observational study seeks to emulate a randomized experiment (\cite{cochran1965planning}, \cite{rosenbaum2010design} Section 1.2) by matching on observed covariates so that any departure of the observed outcomes from the null distribution can be ascribed directly to the treatment. However, in an observational study, failure to control for some relevant, unobserved confounder can undermine the attribution of a causal effect. Thus, a researcher often looks to account for some level of bias due to unmeasured confounding.

\subsection{A model for sensitivity analysis} \label{sens-analysis}

In order to evaluate how the conclusions of a study would change under different levels of unmeasured confounding, a typical strategy is to perform sensitivity analyses of the results. One common framework is the Rosenbaum sensitivity model (\cite{rosenbaum2002observational}), parameterized by a single parameter $\Gamma$, which quantifies the amount of departure of the observed treatment exposure in a matched study from random assignment. A researcher often calculates a set of $p$-values $[\underline{p}_\Gamma,\bar{p}_\Gamma]$ in these analyses, whereas $\underline{p}_1 = \bar{p}_1$. If these $p$-values remain significant at larger values of $\Gamma$, then an analysis is more robust to possible violations of the unconfoundedness assumption. 

The Rosenbaum sensitivity model presupposes that the odds of receiving treatment in a matched set with the same values of $\bm x$ is
\begin{equation} \label{rosen-model} \dfrac 1\Gamma \le \dfrac{\text{pr}(Z_{ij} = 1|\mathcal{F},\mathcal{Z})}{\text{pr}(Z_{ik}=1|\F,\mathcal{Z})}\le \Gamma \quad (1\le j; k\le n_i; 1\le i\le I) %
\end{equation}
with independent assignment across different matched sets. A value of $\Gamma =1$ yields random assignment, whereas any value $\Gamma\ge 1$ implies an unknown but restricted departure from random assignment due to some unobserved covariate. In the subsequent discussion, we focus on the special case of matched pairs; that is, $n_i = 2$. This condition is relaxed in the Supplementary Material. 

Let $Y_i^l = (Z_{i1} - Z_{i2})(R^l_{i1} - R^l_{i2})$ denote the treatment-control difference in the $i$th matched set for the $l$th outcome. A popular choice of statistics for testing the null are the signed score statistics 
\begin{equation}
    \label{pairstats}
    T^l(Z,R) = \dfrac{\sum_{i=1}^ I\sgn(Y_i^l)q^l_i}{\sum_{i=1}^I q^l_i}
\end{equation}
where $\sgn(y) = \1(y\ge 0)$ and $q_i^l$ is some function of $|Y_i^l|$ such that $q_i^l = 0$ if $Y_i^l = 0$. The statistic in Equation \ref{pairstats} has been normalized by $\sum_{i=1}^I q_i$ following \cite{zhao2018sensitivity}, so that it remains confined between zero and one. For instance, $q_i^l = \1(|Y_i^l|>0)$ yields the sign-statistics and $q_i^l = \text{rank}(|Y_i^l|)$ gives the well-known Wilcoxon signed rank statistics. Under the null hypothesis and assuming random treatment assignment (i.e., $\Gamma =1$ in Equation \ref{rosen-model}), conditional on $\F$ and $\mathcal{Z}$, $\sgn(Y_i^l)$ are independent and identically distributed Bernoulli(${1}/{2}$) random variables and $q_i^l$'s are fixed constants, from which the null distribution of the statistics can be derived. However, the null is composite when $\Gamma>1$, so one seeks to maximize the $p$-value subject to Equation \ref{rosen-model}.

\cite{rosenbaum2002observational} showed that one can obtain 
$\underline p^l_{{\Gamma}}:= \text{pr}(T^l_{\underline\Gamma}\ge t|\F,\Z)\le \text{pr}(T^l\ge t|\F,\Z)\le \text{pr}(T^l_{\bar\Gamma}\ge t|\F,\Z)=: \bar p^l_{\Gamma}$,
where $T^l_{\bar\Gamma}$ is the sum of $I$ independent random variables taking the value {$q^l_i/\sum_{j=1}^I q_j^l$} with probability $\Gamma/(1+\Gamma)$ and zero with probability $1/(1+\Gamma)$, whereas $T^l_{\underline\Gamma} = T^l_{\overline{1/\Gamma}}$. A standard sensitivity analysis computes the plausible $p$-values $\underline p^l_{\Gamma}$ and $\bar p^l_{\Gamma}$ for the statistic in Equation \ref{pairstats} under Model \ref{rosen-model} using the previous result. Indeed, for large sample size $I$, $T_{\bar\Gamma}$ satisfies a central-limit theorem  (\cite{sidak1999theory}) when {$\max_i (q_i^l)^2/\sum_{i=1}^I (q_i^l)^2\to 0$}. Conditional on $\F$ and $\Z$, we {thus assume}
\begin{equation}
    \label{clt}
    \dfrac{I^{1/2}\left(T^l_{\bar\Gamma} - \frac{\Gamma}{1+\Gamma}\right)}{{\left\{\frac{\Gamma}{(1+\Gamma)^2}(\sigma^l_q)^2\right\}}^{1/2}}\overset{d}{\to}\mathcal{N}(0,1), \text{ with }(\sigma^l_q)^2 :=\lim_{I\to\infty}(\sigma_{q,I}^l)^2=\lim_{I\to\infty} \dfrac{\sum_{i=1}^n (q^l_i)^2/I}{(\sum_{i=1}^n q^l_i/I)^2} <\infty. %
\end{equation}

Another common practice in observational studies is to report the sensitivity value (\cite{zhao2018sensitivity}), defined as $\Gamma^*_\alpha(\bm Z,\bm R^l) = \inf\{\Gamma >0: \bar p_\Gamma^l (\bm Z,\bm R^l) \ge \alpha\}$.
Simply stated, the sensitivity value is the smallest level of confounding required to render the $p$-value insignificant. While a typical sensitivity analysis is conducted for $\Gamma\ge 1$, the model is well-defined for any positive $\Gamma$ and defining it as before removes the mass at one due to truncation. Before proceeding to the next section, it is important to substantiate the difference between sensitivity value and the more commonly used design sensitivity $\tilde\Gamma^l$. Design sensitivity is defined as the parameter $\tilde\Gamma^l$ such that $H_{0l}$ is rejected under true causal effect and no unmeasured confounding with probability one when $\Gamma<\tilde\Gamma_l$ and with probability zero when $\Gamma>\tilde \Gamma_l$ as $I\to\infty$ (\cite{rosenbaum2004design}). Sensitivity value, on the other hand, is a deterministic function of the data, and is hence a random variable instead of an asymptotic concept. In particular, design sensitivity is the stochastic limit of the sensitivity value statistic as $I\to\infty$ and is insensitive to the sample size.%

\section{Identifying Outcomes More Robust to Hidden Bias} \label{method}
\subsection{Principles of selection and a simple approach}
A researcher may use sample splitting to choose outcomes to test in an observational study. We define the approach of sample splitting for hypothesis screening by the following three steps.
{
    \begin{enumerate}[label={Step \arabic*: }, leftmargin=*]
    \item Randomly partition the dataset into planning and analysis sets, denoted as $\mathcal{G}_{\p}$ and $\mathcal{G}_{\a}$ respectively, such that $\mathcal{G}_{\p} \cap \mathcal{G}_{\a} = \emptyset$.
    \item Choose a subset of outcomes to test $S\subseteq \{1,\cdots, L\}$ from $\G_{\p}$ without looking at $\G_{\a}$. One may similarly choose from the planning sample a set of significance levels $\{\alpha_l\}_{l=1}^L$ such that $\sum_{l\in S}\alpha_l = \alpha$.
    \item Report $\mathcal{T} \subseteq S$, the set of outcomes with null hypotheses rejected on $\mathcal{G}_{\a}$, where each outcome $l$ is tested at level $\alpha_l\1(l\in S)$.
    \end{enumerate}
\begin{proposition}
    \label{validity}
    Sample splitting for hypothesis screening satisfies:
    $$\text{pr}(\exists l\in \{1,\ldots, L\}: \text{$H_{0l}$ is true, $l \in \mathcal{T}$} \text{ 
 }|\text{  }\mathcal{G}_{\p})\le \alpha$$ 
\end{proposition}
\begin{proof}$\text{pr}(\exists l: H_{0l} \text{ is true, $l \in \mathcal{T}$}\text{  }|\text{  }\mathcal{G}_{\p}) \le \sum_{l=1}^L \text{pr}(l \in \mathcal{T}\text{  }|\text{  }\mathcal{G}_{\p}, H_{0l} \text{ is true}) \le \sum_{l=1}^L \alpha_l\1(l\in S) = \alpha.$\end{proof}
\textit{Remark.} Proposition \ref{validity} implies that the family-wise error rate (FWER) of this approach is controlled at level $\alpha$.
}

Although this procedure facilitates valid exploration and hypothesis screening in the planning sample, there is currently no rigorous strategy for choosing outcomes to test using split samples in an observational study (\cite{heller2009split,small2024protocols}). As a guiding principle to establish a systematic approach to outcome selection, we adhere to the advice of \cite{rosenbaum2004design} and prioritize designs with higher design sensitivity, all else being equal. However, design sensitivity is a complicated function of outcome signal strength, which often must be estimated, study design, and choice of test statistics. One approach is to estimate the design sensitivity via the sensitivity value in the planning sample, as this is known to be consistent for design sensitivity (\cite{zhao2018sensitivity}). A researcher could then select those outcomes for which the sensitivity value is larger than control level $\Gamma_{\c}$, which is the level of bias the researcher aims to account for in the analysis sample. A natural implementation is thus to directly test for the outcomes at some $\alpha$ and $\Gamma_{\c}$, which is delineated in Method \ref{one-shot}. %
\begin{method}
    \label{one-shot}
    {(Na\"ive)}. Consider the approach of sample splitting for hypothesis screening. To control for bias at level $\Gamma_{\c}$ in the analysis sample, test for those outcomes $H_{0l}:l\in S^\text{Na\"ive}_{\p}(\Gamma_{\c},\alpha)$, where 
$$S^\text{Na\"ive}_{\p}(\Gamma_{\c},\alpha) = \{l\in\{1,\ldots, L\}: \bar{p}^l_{\Gamma_{\c}}(\bm Z, \bm R^l)\le\alpha\}$$
\end{method}

In some settings, such as observational studies with moderate to large sample sizes and high levels of control for bias due to unmeasured confounding, this strategy can be useful in selecting hypotheses prior to data analysis (\cite{heller2009split}). To illustrate, consider a setup with $N=1000$ subjects randomly assigned to treatment or control with probability $1/2$ and $L=500$ outcomes of potential interest. Suppose that one wishes to maintain considerable control for possible confounding biases in the analysis sample, leading us to set $\Gamma_{\c} = 3.5$. All control potential outcomes $r^l_{C_{ij}}$ are generated independently from the standard Normal distribution and treated potential outcomes are taken as: $r_{T_{ij}}^l = r_{C_{ij}}^l + \mathbbm{1}(l \in \{1,\ldots, 10\})$.
{To be precise, we define power throughout the manuscript as the proportion of non-null hypotheses that are rejected, which is the true positive rate}. Using the Wilcoxon signed rank statistic and a planning sample proportion of $0.20$, the power for the Na\"ive method is approximately $0.24$ while the power for Bonferroni correction without data splitting is around $0.16$ when averaged across 1,000 simulations at $\alpha=0.05$. 

Unfortunately, there also exist numerous situations which do not lend themselves well to this approach for the design of a study. For instance, the Na\"ive method often performs poorly compared to full-sample techniques in settings with smaller samples, fewer hypotheses of potential interest, and when controlling for small values of $\Gamma_{\c}$. Despite its intuitive appeal and concise implementation, these shortcomings warrant the development of a more sophisticated technique to select outcomes and enhance design. The procedure we introduce next will consider the sensitivity value and its estimated variability for each outcome on the planning sample to construct predictive intervals on the analysis sample and render a more robust selection of outcomes.

\subsection{Selection via the sensitivity value} \label{sec:ourmethod}

In a data splitting regime tasked with hypothesis screening, a researcher prioritizes the meaningful filtering of outcomes so that those tested in the analysis sample have a reasonable chance of turning out to be significant. As discussed in Section \ref{sens-analysis}, the sensitivity value is a deterministic function of the data instead of an asymptotic concept like design sensitivity, so we can expect outcome selection using the sensitivity value to be more purposeful in finite samples, especially when working with small or moderate sample sizes.

Under $\Gamma^*_\alpha(\bm Z,\bm R^l)$, we define the transformed sensitivity value $\kappa_\alpha^l(\bm Z,\bm R^l) = \frac{\Gamma^*_\alpha(\bm Z,\bm R^l)}{1+\Gamma^*_\alpha(\bm Z,\bm R^l)}$. In a data splitting framework, we partition $(\bm Z,\bm R) = \cup_{k\in \{\text{planning, analysis}\}}(\bm Z_k,\bm R_k)$ and modify the notation accordingly. Now, suppose that the statistic $T^l$, as defined in Equation \ref{pairstats}, satisfies 
\begin{equation} \label{alt-clt}
\dfrac{I^{1/2}(T^l-\mu_F^l)}{\sigma_F^l}\overset{d}{\to}\mathcal{N}(0,1). 
\end{equation}
This assumption is valid under weak regularity conditions (\cite{hettmansperger1984statistical}). {It is worth noting that the earlier Condition \ref{clt} concerns the bounding random variable $T^l_{\bar\Gamma}$, whose conditional distribution on $\mathcal{F}, \mathcal{Z}$ is data-independent. Only under the null does $T_\Gamma^l$ stochastically dominate the test statistic $T^l$. In contrast, Condition \ref{alt-clt} concerns the distribution of $T^l$ conditional on $\mathcal{F}, \mathcal{Z}$, which depends on the underlying data distribution.}. Using Results \ref{clt} and \ref{alt-clt}, \cite{zhao2018sensitivity} showed that one may obtain 
\begin{equation*} \label{Qing-main}
I^{1/2}(\kappa_\alpha^l - \mu_F^l)\overset{d}{\to}\mathcal{N}\left(-\sigma_q^l\Phi^{-1}(1-\alpha){\left\{\mu_F^l(1-\mu_F^l)\right\}}^{1/2}, (\sigma^l_F)^2 \right),
\end{equation*}
where $\mu_F^l$ is related to Rosenbaum's design sensitivity as $\mu_F^l = \tilde\Gamma^l/(1+\tilde\Gamma^l)$, $\Phi$ denotes the standard Normal distribution function, and $F$ signifies dependence on the distribution of outcomes. %

Suppose we test outcome $l$ in the analysis sample at level $\alpha_l$. We know that outcome $l$ will be deemed significant if $\kappa^l_{\alpha_l}(\bm Z_{\a}, \bm R^l_{\a})> \kappa_{\c}$, where $\kappa_{\c} = \Gamma_{\c}/(1+\Gamma_{\c})$ is the transformed bias level at which the researcher would like to control. If we can create a predictive interval for $\kappa^l_{\alpha_l}(\bm Z_{\a}, \bm R^l_{\a})$, abbreviated as $\kappa^l_{\alpha_l}$, based on the data from the planning sample containing $rI$ matched pairs, $r\in(0,1)$, then we can assess whether that interval lies towards the right of $\kappa_{\c}$ and select the outcomes to test accordingly. {Interestingly, since the sensitivity values of both the planning and the analysis sample are consistent and independent estimators of the same parameter (design sensitivity), they may be combined to create the necessary predictive interval, as we delineate in Theorem \ref{ourmain1}}.

\begin{theorem}
\label{ourmain1}
Assume that {the central limit theorem \ref{clt} for the bounding random variable $T^l_{\bar\Gamma}$ holds, and the test statistic $T^l$ satisfies }\ref{alt-clt}. Then, we have
\begin{align*}
    I^{1/2}\left(\kappa^l_{\alpha_l} - \kappa^l_{\alpha_{\p}}\right) &+
\left \{\kappa_{\alpha_{\p}}^l(1-\kappa_{\alpha_{\p}}^l)\right \}^{1/2}\sigma^l_q\left\{\dfrac{\Phi^{-1}(1-\alpha_{\p})}{r^{1/2}} - \dfrac{\Phi^{-1}(1-\alpha_l)}{(1-r)^{1/2}}\right\} \\
&\overset{d}{\to}\mathcal{N}\left(0,\dfrac{(\sigma^l_F)^2}{r(1-r)}\right).
\end{align*}
\end{theorem}

We need one final detail to derive a confidence set for $\kappa^l_{\alpha_l}$. The astute reader will note that $\sigma^l_F$ is still unknown in Theorem \ref{ourmain1} and cannot be estimated directly since $\kappa^l_{\alpha_{\p}}$ can be only obtained once in the planning sample. To mitigate this issue, we take bootstrap samples of the matched pairs in $\mathcal{G}_{\p}$ and compute the sensitivity values $\{\kappa^l_{\alpha_{\p},b}\}_{b \in \{1, \ldots, B\}}$. We discuss the potential benefits that arise from our use of the bootstrap in Subsection \ref{edgeworth-section}. Then, we can estimate ${(\hat\sigma_F^l)^2} = I \widehat\V(\kappa_{\alpha_{\p},b}^l)$, where $\widehat\V(\cdot)$ denotes the sample variance. Together, this constitutes our Sens-Val Method \ref{bootstrap}.

\begin{method}
    \label{bootstrap}
    {(Sens-Val)}. Consider the approach of sample splitting for hypothesis screening. To control for bias at level $\Gamma_{\c}$ in the analysis sample, for each outcome $l$:
    \begin{itemize}
        \item Compute $\kappa^l_{\alpha_{\p}}$, the sensitivity value for the $l$th outcome at level $\alpha_{\p}$.
        \item Take $B$ bootstrap samples of $I_{\p}$ matched pairs from $(\bm Z_{\p}, R^l_{\p})$ and calculate sensitivity values $\{\kappa^l_{\alpha_{\p},b}\}_{b \in \{1, \ldots, B\}}$. Compute $(\hat\sigma^l_F)^2 = I \widehat\V(\kappa_{\alpha_{\p},b}^l)$.
    \item Test for the outcomes $H_{0l}:l\in S^\text{Sens-Val}_{\p}(\Gamma_{\c}, \alpha_{\p})$, where
    \begin{align*} S^\text{Sens-Val}_{\p}(\Gamma_{\c},\alpha_{\p}) &= \left\{l\in [L]: \kappa^l_{\alpha_{\p}} + \dfrac{{\{\kappa_{\alpha_{\p}}^l(1-\kappa_{\alpha_{\p}}^l)\}}^{1/2}\sigma^l_q}{I^{1/2}}\left\{\dfrac{\Phi^{-1}(1-\alpha_{\p})}{r^{1/2}} \right.\right.\\
    &\left.- \dfrac{\Phi^{-1}(1-\alpha_l)}{(1-r)^{1/2}}\right\}%
    \left. >\kappa_{\c} - \dfrac{\hat\sigma^l_F\Phi^{-1}(1-\alpha_{\cv})}{\{Ir(1-r)\}^{1/2}} \right\}
    \end{align*}
    where $[L] = \{1,\ldots, L\}$, and $\{\alpha_l\}_1^L$ and $\alpha_{\cv}$ are hyperparameters to be determined by the researcher.
    \end{itemize}
\end{method}
For a detailed discussion pertaining to hyperparameters, see the Supplementary Material. We present a corollary to Theorem \ref{ourmain1} below, which follows from the consistency of $\hat\sigma_F^l$.
\begin{corollary} \label{ourcor1}
     If $\hat\sigma^l_F$ is consistent for $\sigma_F^l$, then \begin{align*} \text{pr}\left(\vphantom{\dfrac{\hat\sigma_F^l\Phi^{-1}(1-\alpha_{\cv})}{[r\{1-r\}]^{1/2}}}\kappa^l_{\alpha_l} \right.&\ge \kappa^l_{\alpha_{\p}} + \dfrac{{\{\kappa_{\alpha_{\p}}^l(1-\kappa_{\alpha_{\p}}^l)\}}^{1/2}\sigma^l_q}{I^{1/2}}\left\{\dfrac{\Phi^{-1}(1-\alpha_{\p})}{r^{1/2}} - \dfrac{\Phi^{-1}(1-\alpha_l)}{(1-r)^{1/2}}\right\} \\ &\hspace{12em}\left.+\dfrac{\hat\sigma_F^l\Phi^{-1}(1-\alpha_{\cv})}{\{Ir(1-r)\}^{1/2}} \right)= (1 - \alpha_{\cv}) + o(1) \end{align*}
\end{corollary}

Corollary \ref{ourcor1} provides a $(1-\alpha_{\cv})$ predictive interval for the sensitivity value of the $l$th outcome when tested on the analysis sample at level $\alpha_l$. 
We use one-sided confidence regions in Method \ref{bootstrap} in order to maximize the probability that true causal effects are detected by our method. Since a signal cannot be detected if it is not tested on the analysis sample, we intend to be reasonably cautious in filtering out outcomes during the planning stage of our procedure, the validity of which is ensured by Proposition \ref{validity}.

The Sens-Val method often demonstrates competitive or superior performance compared to the Na\"ive approach and full-sample alternatives. Consider a simple setup with $N=200$ subjects randomly assigned to treatment or control with probability $1/2$ and $L=20$ outcomes of possible interest. Suppose that all control potential outcomes $r^l_{C_{ij}}$ are generated independently from the standard Normal distribution and treated potential outcomes are taken as: $r^l_{T_{ij}} = r^l_{C_{ij}} + 3/4 \cdot \mathbbm{1}(l \in \{1,\ldots, 5\})$.
As before, we consider the Wilcoxon signed rank statistic and a planning sample proportion of $0.20$. While the power for Sens-Val is approximately $0.648$, Bonferroni correction without data splitting is around $0.583$, and the Na\"ive method is approximately $0.275$ when averaged across 1,000 simulations and controlled at levels $\alpha=0.05$ and $\Gamma_{\c} = 1.5$. 

Recall that the aforementioned Wilcoxon signed rank statistic corresponds to \( q_i = \text{rank}(|Y_i|) \). Thus, we know $\sum_i q_i = I(I + 1)/2$ and $\sum_i q_i^2 = I(I + 1)(2I + 1)/6$. By \cite{hettmansperger1984statistical}, we then have $\mu_F = P(Y_1 + Y_2 > 0)$ and $\sigma^2_F = 4\text{  }\{P(Y_1 + Y_2 > 0) - P(Y_1 + Y_2 > 0, \text{  }Y_1 + Y_3 > 0)^2\}.$ In our previous setup, Wilcoxon's test has mean \( \mu_F = \Phi(3 \surd{2}/4) \approx 0.856 \), which corresponds to design sensitivity \(\tilde\Gamma^l = \mu_F/(1 - \mu_F) \approx 5.924\), for $l \in \{1,\dots,5\}$. As we have fixed $\Gamma_{\c} = 1.5$, testing outcomes $l$ with large design sensitivity is equivalent to testing the true signals. Imagine an oracle that tests only these non-null outcomes. The average power of this oracle when applied to the previous setting is roughly $0.659$, which is only slightly higher, if far less realistic, than Sens-Val.

{In the above example we saw that Sens-Val performs better than the Na\"ive method. Proposition \ref{power}, in fact, illustrates that one can always ensure that the performance of Method \ref{bootstrap} is always better than that of Method \ref{one-shot}.}

\begin{proposition} \label{power}
    {For specified number of outcomes $L$, there exists a choice of hyperparameters $\{\alpha_l\}_1^L$ and $r$ based on $\alpha_{\p}$, and $L$ such that the Sens-Val method in Method \ref{bootstrap} with these hyperparameters and any $\alpha_{\cv}\le 0.5$ is guaranteed to select more outcomes than the Na\"ive method in Method \ref{one-shot}.}
\end{proposition}

\section{Theoretical Properties} \label{theory}

\subsection{Edgeworth expansion and unknown parameter estimation} \label{edgeworth-section}

One of the features of Method \ref{bootstrap} is its use of the bootstrap to estimate the unknown parameter $\sigma_F^l$, in order to account for the variability of the sensitivity value. This allows us to construct a predictive interval for the analysis sample sensitivity value based only on the information contained in the planning sample, yet we must rely on the accuracy of the bootstrap procedure in order to do so. We derive a novel Edgeworth expansion for the sensitivity value and discuss how it can be related to a bootstrap distribution leading to a higher order accuracy under standard assumptions, which may be of independent theoretical interest. This new expansion demonstrates its variability up to $O(I^{-1})$, from previously known results of $O(I^{-1/2})$ from \cite{zhao2018sensitivity}, and  %
complements Corollary \ref{ourcor1} by demonstrating the reliability of Method \ref{bootstrap}.

\begin{proposition}
    \label{vble-expand} Suppose the conditions of Equation \ref{clt} and Equation \ref{alt-clt} hold. Additionally, assume $c^l_{q,I} = \dfrac{\frac 1n\sum_{i=1}^I (q^l_i)^3}{\left(\frac 1n\sum_{i=1}^n q^l_i\right)^3}$ is such that $c_q^l = \lim_{I\to\infty} c^l_{q,I} <\infty$. Define $T_{\text{alt}}^l = \dfrac{I^{1/2}(T^l - \mu_F^l)}{\sigma_F^l}$. Then,
    \begin{align*}
        V^l := \dfrac{I^{1/2}(\kappa_\alpha^l - \mu_F^l)}{\sigma_F^l} + \dfrac{z_\alpha\sigma^l_{q,I}}{\sigma_F^l}\left\{\kappa_\alpha^l(1-\kappa_\alpha^l)\right\}^{1/2} = T_{\text{alt}}^l - (2\mu_F^l-1)\dfrac{c_1}{I^{1/2}}+ O(I^{-1}),
    \end{align*}
    where $c_1 = c_{q,I}^lg(z_\alpha)/\sigma_F^l$ and $g(x) = (x^2-1)/6$. 
\end{proposition}

Proposition \ref{vble-expand} allows us to consider the Edgeworth expansions for $V^l$ based on that for $T_{\text{alt}}^l$, assuming that one exists as in Equation \ref{edge-ass} below:
\begin{equation}
    \label{edge-ass}
    \text{pr}(T_{\text{alt}}^l\le x) = \Phi(x) + \dfrac{\phi(x)b^l(x,F)}{I^{1/2}} + O(I^{-1}) \text{ and }\text{pr}_*(T_{\text{alt}}^l\le x) = \Phi(x) + \dfrac{b^l(x,F_I)}{I^{1/2}} + O(I^{-1}),
\end{equation}
where $b(\cdot,F)$ is a polynomial in $x$ that depends on the parameters of the distribution $F$, $F_I$ denotes the empirical distribution of $T_{\text{alt}}^l$, and $\Phi$ and $\phi$ denote the standard Normal distribution function and density, respectively. Expansions of this form have been derived and discussed in \cite{albers1976asymptotic}, where the coefficients of $b^l$ are generally associated with skewness of the distribution of $T_{\text{alt}}^l$.

\begin{theorem}
    \label{edgeworth}
    Suppose the conditions of Proposition \ref{vble-expand} and Equation \ref{edge-ass} hold. Then, 
    \begin{align*}
        \text{pr}(V^l\le v)= \Phi(v) + \dfrac{1}{I^{1/2}}\phi(v)\left\{(2\mu_F^l-1)c_1 + b^l(v,F)\right\} + O(I^{-1})
    \end{align*}
    Analogous expressions hold for the bootstrapped version with $\text{pr}$ replaced by $\text{pr}_*$, $\mu_F$ and $F$ replaced by $\kappa_\alpha$ and empirical distribution $\hat F$, respectively.
\end{theorem}

While Theorem \ref{edgeworth} produces an expansion of the distribution function of the bias-corrected version of $\kappa_\alpha^l$, a similar expansion is also derived for the uncorrected version in the Supplementary Material. Theorem \ref{edgeworth} reveals that the difference between the distribution for $V^l$ and its bootstrapped version is $O(I^{-1})$. This is because %
$$\text{pr}(V^l\le v) - \text{pr}_*(V^l\le v) = \dfrac{1}{I^{1/2}}\left\{2c_1(\mu_F^l - \kappa_\alpha) + b^l(v,F) - b^l(v,\hat F)\right\} + O(I^{-1}),%
$$
of which $\mu_F^l - \kappa_\alpha$ and $b^l(v,F) - b^l(v,\widehat{F})$ are both $O(I^{-1/2})$, rendering the approximation $O(I^{-1})$ consistent. This stands in contrast to the central limit theorem approximation, which only provides accuracy up to $O(I^{-1/2})$. Therefore, not only does the bootstrap-based procedure work as intended, but it can also bring about enhanced accuracy compared to the usual Normal approximation. Although Method \ref{bootstrap} is described utilizing the Normal approximation combined with the bootstrap estimate of $\sigma_F^l$—a choice driven by its already adequate performance and to help ensure clarity on part of the reader—using the quantiles of the bootstrap distribution directly could further improve accuracy. This idea is elaborated upon in Section \ref{discussion}.

Theorem \ref{edgeworth} also allows us to consider the accuracy of the bootstrap distribution in general and that of its moments. \cite{wasserman2006all} discusses how the accuracy of the bootstrap of a functional relates to its Hadamard differentiability and one can observe that Method \ref{bootstrap} is essentially a version of the non-parametric delta method. With respect to estimation of moments like $\sigma_F^l$, \cite{bickel1981some} uses the concept of von-Mises functionals to demonstrate the accuracy of plug-in bootstrap U-statistic-based functionals. Indeed, since $(\sigma^l_F)^2$ is the asymptotic limit of $E[\{\kappa_\alpha^l - (\kappa_\alpha^l)'\}^2]/2$, where $\kappa_\alpha^l$ and $(\kappa_\alpha^l)'$ are \textit{iid} copies of the transformed sensitivity values, the discussion in Section 2 of \cite{bickel1981some} allows us to infer consistency of $\sigma_{F_{\text{boot}}}^l$.

\subsection{Local asymptotic power} \label{local-qual}

We assess the quality of the filter outlined by Method \ref{bootstrap} in terms of the probability that it selects a non-null outcome when its corresponding signal is small. We demonstrate that our filter is powerful in selecting important outcomes under local alternatives, thereby suggesting that even if the strength of true signals is barely above the detection threshold, Method \ref{bootstrap} has a high chance of selecting these outcomes.%

We consider a location family such that the pairwise difference in the treated-control pair is obtained by a distribution from the family $\mathcal{F}^l = \{F^l_\theta: F^l_\theta \text{ admits a density of form }f_l(\cdot -\theta)\}$, where $\theta$ quantifies the treatment effect. We denote $\mu^l_{F_\theta}$ as $\mu(\theta)$ and the same for $\sigma(\theta)$, suppressing the $l$ superscript as it would be clear from the context. We take $\theta_0$ such that $\mu(\theta_0) = \tilde\Gamma/ (1+\tilde\Gamma)$ and $\mu(\cdot)$ to be differentiable at $\theta_0$. To consider the effect of moderate sample sizes without the asymptotic degeneracy with sample size, we consider the class of local alternatives $\theta_I = \theta_0 + {h}/{\surd{I_{\p}}} = \theta_0 + {h}/{(r\surd{I})}$. Although we typically do not expect a true effect to depend on the sample size, this exercise allows us to control the degeneracy of large samples. Suppose \ref{alt-clt} also holds under this local alternative with $\theta_I$ given as above. This is not directly implied by \ref{alt-clt}, as now the distribution parameters also change with the sample size $I$. Nonetheless, the local version of \ref{alt-clt} can hold under contiguity arguments, which we discuss in Proposition \ref{le-cam}. This version also allows us to obtain Theorem \ref{localasympot}.
\begin{proposition}
    \label{le-cam}
    Suppose the statistic $T^l$ in Equation \ref{pairstats} be such that $q_i^l$ is generated by $q_i^l = \psi(\text{rank}{(|Y_i^l|)}/(I+1))$ for some differentiable function $\psi$ on $[0,1]$, such that $0<\int_0^1 \psi(u)\,du <\infty $, $\int_0^1\psi(u)^2<\infty$ and $\psi'$ is bounded. Also, let $F_\theta^l\in \mathcal{F}^l$ is in a location family that admits a positive continuously differentiable density. Then, Equation \ref{alt-clt} can be strengthened to obtain,%
    \begin{equation} \dfrac{I^{1/2}(T^l - \mu^l(\theta_I))}{\sigma^l(\theta_I)}\overset{d}{\to}\mathcal{N}(0,1)%
    \label{st-altclt}\end{equation}
    with $\theta_I = \theta_0 + h/\surd{I}$, $h>0$.
\end{proposition}

\begin{theorem}
\label{localasympot}
Suppose that Equation \ref{st-altclt} holds and $\mu(\theta)$, $\sigma(\theta)$ are differentiable and continuous, respectively, at point $\theta_0$. Then, assuming $\theta_I = \theta_0 + {h}/{\surd{I_{\p}}}$, where $\theta_0$ is such that $\mu(\theta_0) = {\Gamma_{\c}}/({1+\Gamma_{\c}})$, the local power, i.e., the probability that a non-null outcome with location parameter $\theta_I$ is selected via Method \ref{bootstrap}, converges to
$$1-\Phi\left(-\dfrac{z_{\alpha_{\cv}}}{(1-r)^{1/2}}-h\dfrac{(\mu^l)'(\theta_0)}{\sigma^l(\theta_0)}+{z_{\alpha_{l}}}\dfrac{{\Gamma_{\c}}^{1/2}\sigma_q}{(1+\Gamma_{\c})\sigma(\theta_0)}\dfrac{r^{1/2}}{(1-r)^{1/2}}\right).%
$$
\end{theorem}

Theorem \ref{localasympot} reveals that Method \ref{bootstrap} is  powerful even under local alternatives. For instance, let us consider the setting where we use the Wilcoxon signed rank statistic and control for potential unmeasured confounding bias at level $\Gamma_{\c} = 2$, implying a transformed design sensitivity of $2/3$. Under the sequence of alternatives given as $\theta = \Phi^{-1}(2/3)/{\surd{2}} + h/{\surd{I_{\p}}}$, where $h = 0.20$ and $r=0.20$, Method \ref{bootstrap} would have a local power of {roughly $0.97$ for varying choices of $\alpha_l$}, based on the calculations of $\mu(\theta)$ and $\sigma(\theta)$ given by \cite{zhao2018sensitivity}.

While the second term in the expression given by Theorem \ref{localasympot} suggests possible type I error inflation in the planning stage, we emphasize that the primary objective in constructing our filter is establishing its considerable power. Our procedure is formulated intentionally to diminish the probability that true signals are removed prematurely from the pipeline, while still being able to reduce the number of outcomes examined in the analysis stage effectively. Since the validity of our entire procedure is ensured by Proposition \ref{validity} and Corollary \ref{ourcor1} guarantees asymptotic coverage for the sensitivity values in the analysis sample, the substantial power of Method \ref{bootstrap} under local alternatives motivates it as our recommended choice when sample splitting.

\section{Simulation Studies} \label{results}
We conduct simulation studies to assess the performance of the Sens-Val procedure (Method \ref{bootstrap}). We also implement an {oracle that tests only the non-null outcomes on the analysis sample}, the Na\"ive approach (Method \ref{one-shot}) and standard Bonferroni correction for purpose of comparison. These experiments are intended to examine the efficacy of our procedure in both large samples and for our data application. In particular, we emulate the setting proposed by \cite{yadlowsky2022bounds} and show the potential of Sens-Val compared to alternatives.  We then assess the effectiveness of each method with simulated data that has the same dimensions as the Bangladesh flooding dataset.%

The purpose of our first simulation study is to assess the performance of our sensitivity value-based split-sampling framework compared to a Bonferroni correction alternative that does not partition the data. We take a total number of subjects $N = 5000$ and number of outcomes $L = 250$, from which we assume there are five outcomes that constitute ``true signals'' in our data. We choose $d = 20$ covariates and generate $X \sim \text{Uniform}[0, 1]^d$. Then, conditional on $X = x$, we take $U \sim \mathcal{N}(0, \{1 + \sin (3x_1)/2\}^2)$. 
Amongst our five true signals, we set our potential outcomes $r^l_{C_{ij}} = \beta^\top x + \varepsilon$ and $r^l_{T_{ij}} = \tau + r^l_{C_{ij}}$,
while the remaining potential outcomes are generated as:
$r^l_{C_{ij}} = r^l_{T_{ij}} = \beta^\top x + U + \varepsilon,$
where $\beta \sim \mathcal{N}(1,1)$, $\mu \sim \mathcal{N}(0,1)$, and $\varepsilon \sim \mathcal{N}(0,{1}/{4})$ are independent.

Seeking to incorporate the variability characteristic of observational data, we draw the proportion of subjects who are assigned treatment from $\text{Beta}(10,10)$. We then draw the treatment assignment according to $Z \sim \text{Bernoulli}\left(\frac{\exp\{\alpha_0 + x^\top \mu - \log(\Gamma_{\text{data}})1\{u > 0\}\}}{1 + \exp\{\alpha_0 + x^\top \mu - \log(\Gamma_{\text{data}})1\{u > 0\}\}}\right),$
where $\alpha_0$ is a constant to calibrate the overall treatment assignment ratio close to the desired proportion. For a valid sensitivity analysis, we must have $\Gamma_{\c} \geq \Gamma_{\text{data}}$. We consider two setups, one when there is no unmeasured confounding (NUC) in any of the outcomes, unknown to the researcher, where we set $\Gamma_{\text{data}} = 1$, and the other where only the null outcomes suffer from unmeasured confounding (UC) but not the non-null outcomes, setting $\Gamma_{\text{data}} = \Gamma_{\c}$. Even in the latter situation, we do not include unmeasured confounding in the non-null potential outcomes, akin to the Rosenbaum favorable setting (\cite{rosenbaum2010design}, Section 14.2). 
Propensity score pair matching with a caliper is implemented to balance covariate distributions between treated and control units. $20 \%$ of our total matched pairs are randomly selected to be used in the planning sample, while the remaining data are assigned to the analysis sample. We vary both $\tau$ and $\Gamma_{\c}$ across experiments to evaluate the performance of the methods, all of which consider a possible positive treatment effect for each outcome $l$. All experiments take $250$ bootstrap samples and the average power over all the non-null outcomes across $1000$ simulations to produce a univariate measurement of power.

\begin{figure}[t]
\centering
\includegraphics[width = \textwidth-4em]{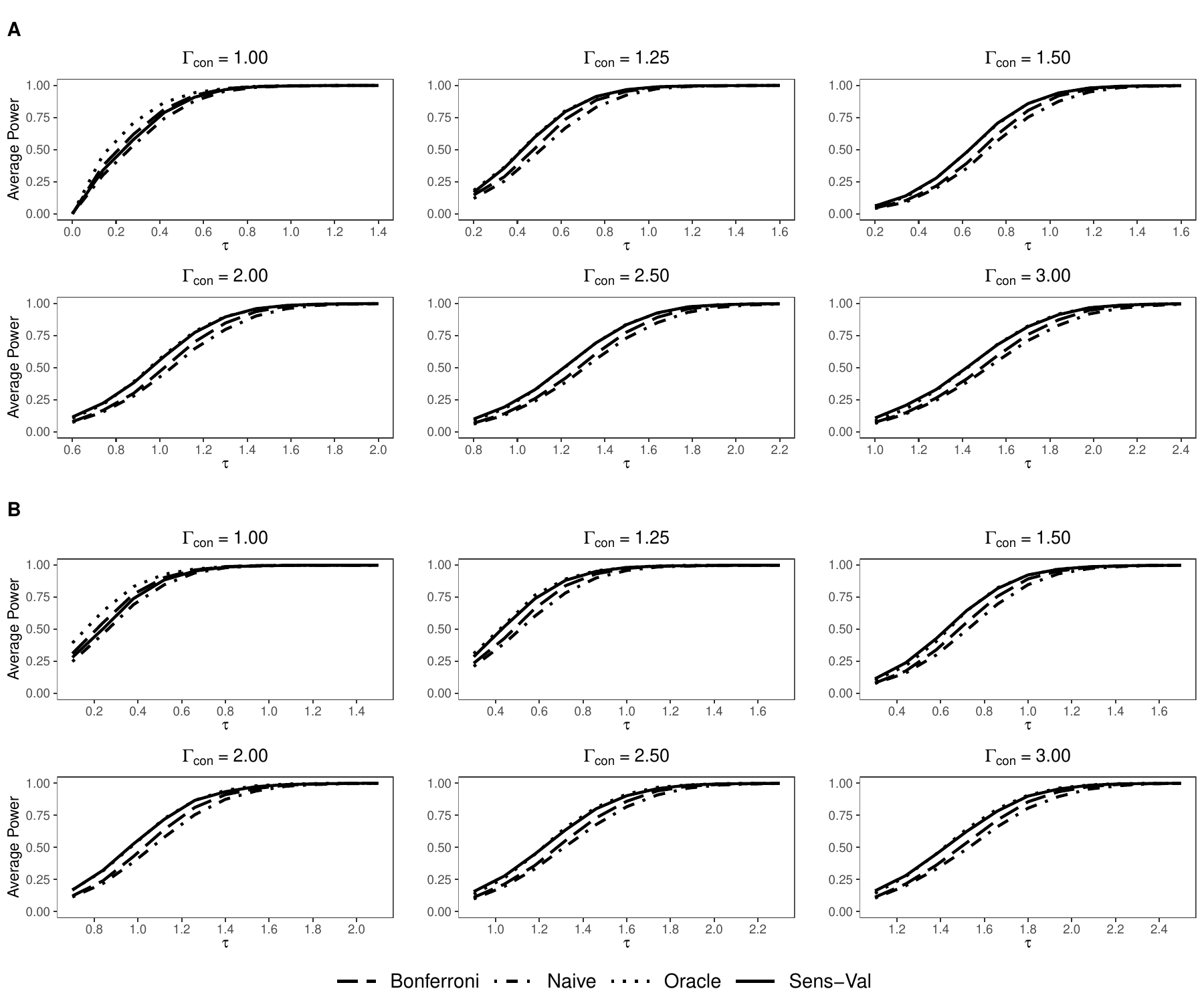}
\caption{{Simulation results in the large sample setting. {(A)} Results for the NUC setting. {(B)} Results for the UC setting.}}
\label{fig:largesample}
\end{figure}

In Figure \ref{fig:largesample}, we display plots showing the results of Sens-Val, the Na\"ive approach, oracle, and Bonferroni correction in the NUC and UC settings at multiple values of $\Gamma_{\c}$. We observe that Sens-Val demonstrates competitive performance across values of $\Gamma_{\c}$ and $\tau$, with increasing benefits at increasing levels of wariness for potential unmeasured confounding biases.
In the second set of experiments, we investigate the behavior of the discussed methods on simulated data with the same dimensions as the Bangladesh flooding dataset. To match the real data, we take a total number of $N = 757$ subjects, $d = 33$ covariates, and $L = 93$ outcomes of interest, from which we assume there are five outcomes that constitute ``true signals''. The remaining variables are generated the same as the previous simulation. In particular, unmeasured confounding $U$, potential outcomes, and treatment assignment are taken the same as in the large-sample setting. However, the proportion of subjects who are assigned treatment is taken equal to $0.71$, consistent with the actual ratio of treated to control subjects in the data.

\begin{figure}[t]
\centering
\includegraphics[width = \textwidth-4em]{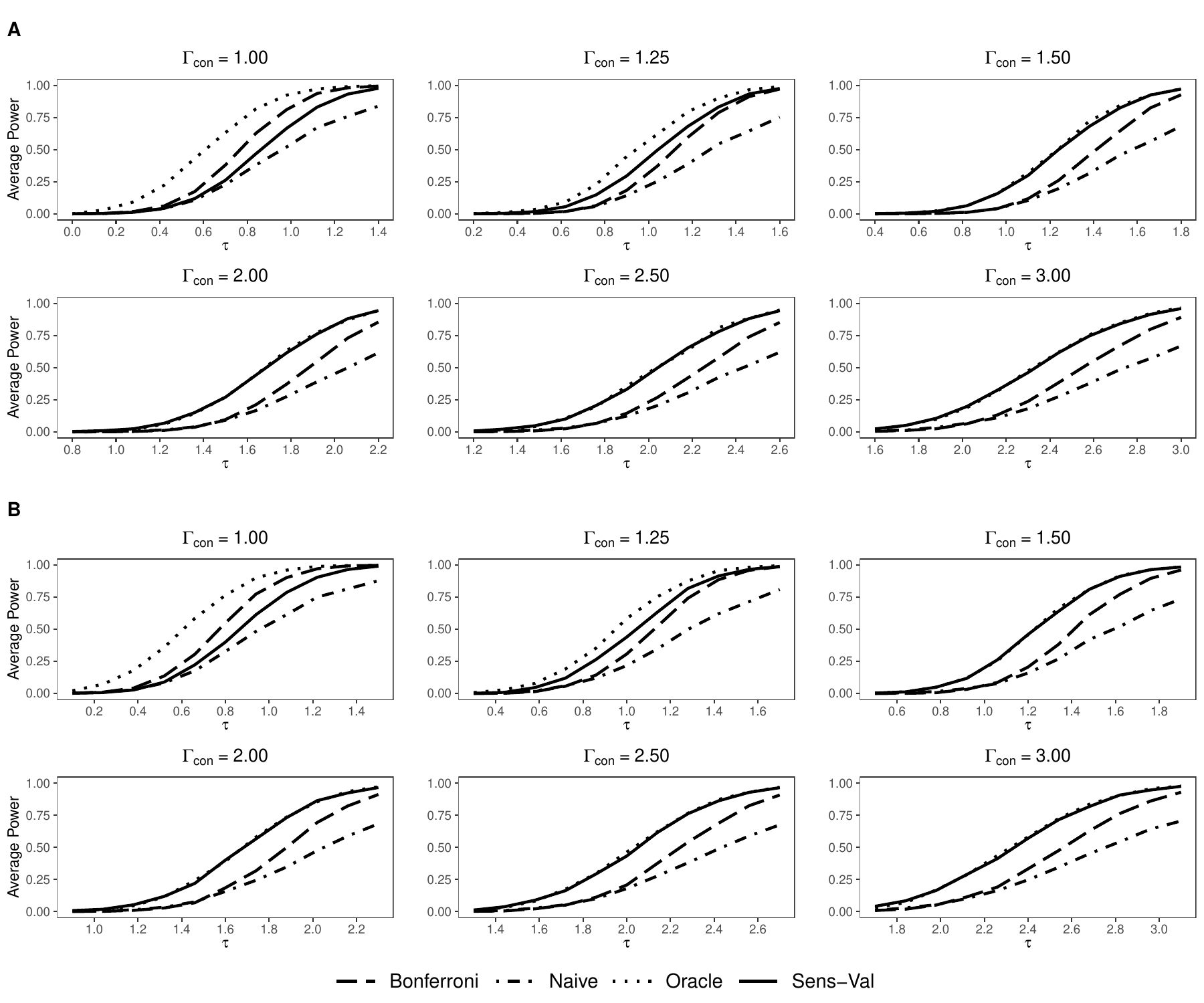}
\caption{{Simulation results in the data-inspired setting. {(A)} Results for the NUC setting. {(B)} Results for the UC setting.}}
\label{fig:datainsp}
\end{figure}

In Figure \ref{fig:datainsp}, we display plots showing the results of Sens-Val, the Na\"ive approach, oracle, and Bonferroni correction in the NUC and UC settings at multiple values of $\Gamma_{\c}$. We observe that, despite maintaining just $80\%$ of the sample for inference, Sens-Val demonstrates competitive or superior power across values of $\Gamma_{\c}$ and $\tau$, with increasing benefits at increasing levels of wariness for potential unmeasured confounding biases. {In particular, we see that the performance of Sens-Val is very close to that of the oracle, indicating that our screening filter is often able to successfully distinguish the set of non-null outcomes from the remaining outcomes of potential interest.} These results show that the Sens-Val procedure can be quite useful for data of moderate sample size and suggest that it should demonstrate strong performance when applied to study the effects of the Bangladesh floods.  
\section{Application to the Impacts of Flooding in Bangladesh} \label{sec:application}

\subsection{Data description and preliminary examinations} \label{data-exam}

Finally, we apply each of the methods to the Bangladesh flooding dataset. In the study by \cite{del20011998}, a detailed household survey was conducted among 757 rural households in seven flood-affected \textit{thanas}. The data was collected at three points in time over a period of a year between November 1998 and December 1999. We focus our analysis on the immediate impacts of the flood on food consumption, food security, health, and nutrition at the household level. Using the first round of household-level data, which was collected between the third week of November and the third week of December 1998, we gather covariates related to household composition, education, assets, and previous flood exposures in order to identify and match similar households. Households and \textit{thanas} were classified by the Water Board to be not affected, moderately affected, severely affected, or very severely affected by the floods, depending on the level and depth of the floodwater. We thus define our treatment to be any of moderate, severe, or very severe flood exposures, rendering our controls to be those matched non-exposed households. We conduct propensity score pair matching with a caliper and obtain satisfactory covariate balance, as shown in the Supplementary Material. This yields $215$ matched pairs to comprise our sample. We fix our planning sample proportion equal to $0.20$ and partition the data into planning and analysis samples. Seeking to investigate the effects of the flood, we collect flood-specific outcomes, as well as those related to food consumption and food security, gender discrimination, illness, nutritional status of children and women, and economic status that were examined by \cite{del20011998}.

Akin to previous observational studies such as \cite{zhang2011using}, statisticians on our team conducted a comprehensive examination of the planning sample and followed up directly with the subject matter expert in the group (del Ninno) to discuss findings and formulate an analysis plan that reflected both his domain knowledge and our statistical judgement. For example, we initially considered outcomes related to both illness and nutritional status in the first round of data, which was collected shortly after flooding affected the region. However, the subject matter expert advised that malnutrition and growth stunting is often due to repeated illnesses and may not be detectable so soon following a disaster. To supplement his guidance, we found on the planning sample little discrepancy between matched households related to the presence of common forms of under-nutrition. On the other hand, we found credible evidence that the intensity of disease, as measured by its duration with correction for extreme values (\cite{bhargava1994modelling}), may have increased with flood exposure. This motivated us to continue to analyze the impacts of the flood on illness in the first round of household-level data, while turning our focus to the third round of data, collected one year after the first round, to assess its long-term effects on nutrition. Moreover, the planning sample also revealed that several of the binary outcomes that report forms of under-nutrition are very sparse and likely too rare to study even with the much larger analysis sample. Accordingly, we restrict our analysis to only include the variables used to construct them, namely body mass index, height-for-age, weight-for-height, and weight-for-age z-scores. Removing variables with missing proportions of at least $75 \%$, we obtain a total of $93$ relevant outcomes, details of which are discussed in the Supplementary Material.

\subsection{Data analysis}

We run our Sens-Val procedure and compare it to the previous alternatives at multiple levels of control for potential unmeasured confounding $\Gamma_{\c}$.  Both sample splitting-based methods exploit the planning sample to estimate the sign of the average treatment effect for each outcome of interest, which is then used to formulate an alternative hypothesis and test the significance of that outcome. Since this is done only using the planning sample, this is a valid strategy which can supplement or supplant domain knowledge when desired. As standard Bonferroni correction does not enjoy the benefits of a planning sample, it conducts two-sided hypothesis tests for each outcome of interest. For binary outcomes, sensitivity analyses and sensitivity value calculations are conducted using McNemar's test statistic (\cite{rosenbaum1987sensitivity,hsu2013calibrating}), while remaining outcomes rely on Wilcoxon's signed-rank statistic with a Normal approximation (\cite{rosenbaum2010design}). We go through each outcome and, if either household in a matched pair has a missing value for that particular outcome, that matched pair is omitted from the sensitivity analyses. This produces valid inference under the assumption that the potential outcomes and treatment assignment are independent of the missingness mechanism conditional on the covariates (MAR; \cite{rubin1976inference}).

We set our $\alpha$-level equal to $0.10$ for sake of comparison with the results obtained by \cite{del20011998}. The previous work uncovered some notable trends in its explorations of the data, such as a rise in food prices, illness, and credit usage, especially for food, and decreases in food production among flood-affected households. However, the authors focused primarily on discovery and did not seek to control the FWER of their analyses. We intend to provide confirmatory evidence for these findings while accounting for the testing of multiple hypotheses in the form of multiplicity correction. We vary $\Gamma_{\c}$ over the values in the set \{1.00, 1.25, 1.50, 2.00, 2.50, 3.00, 4.00, 6.00\} and present our results in Table \ref{tab:results}. A more detailed description of the outcomes we reject is provided in the Supplementary Material. 

\begin{table}[!t]
\centering
\caption{Application of proposed methods to specific household-level effects of 1998 Bangladesh floods. Cells marked with ``\checkmark'' denote that the particular outcome in the row was rejected by the method in the column at the given level of $\Gamma_{\c}$. The shaded column reflects Sens-Val Method \ref{bootstrap}.}
\label{tab:results}
\scriptsize
\setlength{\tabcolsep}{5pt}
\renewcommand{\arraystretch}{1.2}
\resizebox{\textwidth}{!}{\begin{tabular}{lcccccccccccc}
\toprule
& \multicolumn{3}{c}{{$\Gamma_\text{con} = $ 1.00}} & \multicolumn{3}{c}{{$\Gamma_\text{con} = $ 1.25}} & \multicolumn{3}{c}{{$\Gamma_\text{con} = $ 1.50}} & \multicolumn{3}{c}{{$\Gamma_\text{con} = $ 2.00}} \\
\cmidrule(lr){2-4} \cmidrule(lr){5-7} \cmidrule(lr){8-10} \cmidrule(l){11-13}
{Method} & {Bonf.} & {Na\"ive} & \cellcolor{gray!25}{S-V} & {Bonf.} & {Na\"ive} & \cellcolor{gray!25}{S-V} & {Bonf.} & {Na\"ive} & \cellcolor{gray!25}{S-V} & {Bonf.} & {Na\"ive} & \cellcolor{gray!25}{S-V} \\
{No. Tested} & {93} & {27} & \cellcolor{gray!25}{82} & {93} & {15} & \cellcolor{gray!25}{22} & {93} & {10} & \cellcolor{gray!25}{13} & {93} & {6} & \cellcolor{gray!25}{7} \\
\midrule
ft.water & $\checkmark$ & $\checkmark$ & \cellcolor{gray!25}$\checkmark$ & $\checkmark$ & $\checkmark$ & \cellcolor{gray!25}$\checkmark$ & $\checkmark$ & $\checkmark$ & \cellcolor{gray!25}$\checkmark$ & $\checkmark$ & $\checkmark$ & \cellcolor{gray!25}$\checkmark$ \\
repair.cost & $\checkmark$ & $\checkmark$ & \cellcolor{gray!25}$\checkmark$ & $\checkmark$ & $\checkmark$ & \cellcolor{gray!25}$\checkmark$ & $\checkmark$ & $\checkmark$ & \cellcolor{gray!25}$\checkmark$& $\checkmark$ & $\checkmark$ & \cellcolor{gray!25}$\checkmark$ \\
days.water & $\checkmark$ & $\checkmark$ & \cellcolor{gray!25}$\checkmark$ & $\checkmark$ & $\checkmark$ & \cellcolor{gray!25}$\checkmark$ & $\checkmark$ & $\checkmark$ & \cellcolor{gray!25}$\checkmark$& $\checkmark$ & $\checkmark$ & \cellcolor{gray!25}$\checkmark$ \\
days.away.home & $\checkmark$ & $\checkmark$ & \cellcolor{gray!25}$\checkmark$ & $\checkmark$ & $\checkmark$ & \cellcolor{gray!25}$\checkmark$ & $\checkmark$ & $\checkmark$ & \cellcolor{gray!25}$\checkmark$& $\checkmark$ & $\checkmark$ & \cellcolor{gray!25}$\checkmark$ \\
tubewell.cooking & $\checkmark$ & $\checkmark$ & \cellcolor{gray!25}$\checkmark$ & $\checkmark$ & $\checkmark$ & \cellcolor{gray!25}$\checkmark$ & $\checkmark$ & $\checkmark$ & \cellcolor{gray!25}$\checkmark$ & $\checkmark$ & $\checkmark$ & \cellcolor{gray!25}$\checkmark$ \\
tubewell.washing & $\checkmark$ & $\checkmark$ & \cellcolor{gray!25}$\checkmark$ & $\checkmark$ & $\checkmark$ & \cellcolor{gray!25}$\checkmark$ &  &  & \cellcolor{gray!25}$\checkmark$&  &  & \cellcolor{gray!25} \\
food.price.idx & $\checkmark$ & $\checkmark$ & \cellcolor{gray!25}$\checkmark$ & $\checkmark$ & $\checkmark$ & \cellcolor{gray!25}$\checkmark$ &  & $\checkmark$ & \cellcolor{gray!25}$\checkmark$&  &  & \cellcolor{gray!25} \\
food.credit.val.pc & $\checkmark$ & $\checkmark$ & \cellcolor{gray!25}$\checkmark$ & $\checkmark$ & $\checkmark$ & \cellcolor{gray!25}$\checkmark$ &  & $\checkmark$ & \cellcolor{gray!25}$\checkmark$&  &  & \cellcolor{gray!25} \\
credit.val & $\checkmark$ & $\checkmark$ & \cellcolor{gray!25}$\checkmark$ & $\checkmark$ & $\checkmark$ & \cellcolor{gray!25}$\checkmark$ &  &  & \cellcolor{gray!25}&  &  & \cellcolor{gray!25} \\
days.ill & $\checkmark$ & $\checkmark$ & \cellcolor{gray!25}$\checkmark$ &  & $\checkmark$ & \cellcolor{gray!25}$\checkmark$ &  &  & \cellcolor{gray!25}&  &  & \cellcolor{gray!25} \\
pulses.budget & $\checkmark$ & $\checkmark$ & \cellcolor{gray!25}$\checkmark$ &  &  & \cellcolor{gray!25} &  &  & \cellcolor{gray!25}&  &  & \cellcolor{gray!25} \\
rice.consumed.pc & $\checkmark$ &  & \cellcolor{gray!25}$\checkmark$ &  &  & \cellcolor{gray!25} &  &  & \cellcolor{gray!25}&  &  & \cellcolor{gray!25} \\
food.produced.val & $\checkmark$ &  & \cellcolor{gray!25}$\checkmark$ &  &  & \cellcolor{gray!25} &  &  & \cellcolor{gray!25}&  &  & \cellcolor{gray!25} \\
food.produced.val.pc &  &  & \cellcolor{gray!25}$\checkmark$ &  &  & \cellcolor{gray!25} &  &  & \cellcolor{gray!25}&  &  & \cellcolor{gray!25} \\
\bottomrule
\end{tabular}}

\resizebox{\textwidth}{!}{\begin{tabular}{lcccccccccccc}
\toprule
& \multicolumn{3}{c}{{$\Gamma_\text{con} = $ 2.50}} & \multicolumn{3}{c}{{$\Gamma_\text{con} = $ 3.00}} & \multicolumn{3}{c}{{$\Gamma_\text{con} = $ 4.00}} & \multicolumn{3}{c}{{$\Gamma_\text{con} = $ 6.00}} \\
\cmidrule(lr){2-4} \cmidrule(lr){5-7} \cmidrule(lr){8-10} \cmidrule(l){11-13}
{Method} & {Bonf.} & {Na\"ive} & \cellcolor{gray!25}{S-V} & {Bonf.} & {Na\"ive} & \cellcolor{gray!25}{S-V} & {Bonf.} & {Na\"ive} & \cellcolor{gray!25}{S-V} & {Bonf.} & {Na\"ive} & \cellcolor{gray!25}{S-V} \\
{No. Tested} & {93} & {4} & \cellcolor{gray!25}{6} & {93} & {4} & \cellcolor{gray!25}{4} & {93} & {4} & \cellcolor{gray!25}{4} & {93} & {4} & \cellcolor{gray!25}{4} \\
\midrule
ft.water & $\checkmark$ & $\checkmark$ & \cellcolor{gray!25}$\checkmark$ & $\checkmark$ & $\checkmark$ & \cellcolor{gray!25}$\checkmark$ & $\checkmark$ & $\checkmark$ & \cellcolor{gray!25}$\checkmark$ & $\checkmark$ & $\checkmark$ & \cellcolor{gray!25}$\checkmark$ \\
repair.cost & $\checkmark$ & $\checkmark$ & \cellcolor{gray!25}$\checkmark$ & $\checkmark$ & $\checkmark$ & \cellcolor{gray!25}$\checkmark$ & $\checkmark$ & $\checkmark$ & \cellcolor{gray!25}$\checkmark$& $\checkmark$ & $\checkmark$ & \cellcolor{gray!25}$\checkmark$ \\
days.water & $\checkmark$ & $\checkmark$ & \cellcolor{gray!25}$\checkmark$ & $\checkmark$ & $\checkmark$ & \cellcolor{gray!25}$\checkmark$ & $\checkmark$ & $\checkmark$ & \cellcolor{gray!25}$\checkmark$& $\checkmark$ & $\checkmark$ & \cellcolor{gray!25}$\checkmark$ \\
days.away.home & $\checkmark$ & $\checkmark$ & \cellcolor{gray!25}$\checkmark$ & $\checkmark$ & $\checkmark$ & \cellcolor{gray!25}$\checkmark$ & $\checkmark$ & $\checkmark$ & \cellcolor{gray!25}$\checkmark$&  & $\checkmark$ & \cellcolor{gray!25}$\checkmark$ \\
tubewell.cooking &  &  & \cellcolor{gray!25}$\checkmark$ &  &  & \cellcolor{gray!25} &  &  & \cellcolor{gray!25}&  &  & \cellcolor{gray!25} \\
tubewell.washing &  &  & \cellcolor{gray!25} &  &  & \cellcolor{gray!25} &  &  & \cellcolor{gray!25}&  &  & \cellcolor{gray!25} \\
food.price.idx &  &  & \cellcolor{gray!25} &  &  & \cellcolor{gray!25} &  &  & \cellcolor{gray!25}&  &  & \cellcolor{gray!25} \\
food.credit.val.pc &  &  & \cellcolor{gray!25} &  &  & \cellcolor{gray!25} &  &  & \cellcolor{gray!25}&  &  & \cellcolor{gray!25} \\
credit.val &  &  & \cellcolor{gray!25} &  &  & \cellcolor{gray!25} &  &  & \cellcolor{gray!25}&  &  & \cellcolor{gray!25} \\
days.ill &  &  & \cellcolor{gray!25} &  &  & \cellcolor{gray!25} &  &  & \cellcolor{gray!25}&  &  & \cellcolor{gray!25} \\
pulses.budget &  &  & \cellcolor{gray!25} &  &  & \cellcolor{gray!25} &  &  & \cellcolor{gray!25}&  &  & \cellcolor{gray!25} \\
rice.consumed.pc &  &  & \cellcolor{gray!25} &  &  & \cellcolor{gray!25} &  &  & \cellcolor{gray!25}&  &  & \cellcolor{gray!25} \\
food.produced.val &  &  & \cellcolor{gray!25} &  &  & \cellcolor{gray!25} &  &  & \cellcolor{gray!25}&  &  & \cellcolor{gray!25} \\
food.produced.val.pc &  &  & \cellcolor{gray!25} &  &  & \cellcolor{gray!25} &  &  & \cellcolor{gray!25}&  &  & \cellcolor{gray!25} \\
\bottomrule
\end{tabular}
}

\end{table}

As expected, all procedures reject the largest number of outcomes at small levels of $\Gamma_{\c}$ and the fewest number at larger levels of $\Gamma_{\c}$. At $\Gamma_{\c}=1$, our results are very similar to those identified by \cite{del20011998}. In particular, all three procedures identify significant differences in credit.val and food.credit.val.pc, between exposed and non-exposed households. As discussed in the report, borrowing was one of the primary coping mechanisms employed by households in response to the floods. While the proportion of households that took out any loan increased substantially during the flood period, this was especially pronounced for the purchasing of food. Specifically, the percentage of households who took out a loan for the purpose of buying food was only $7 \%$ before the flooding, but increased to almost $16 \%$ by the final month of flooding. This rise in borrowing can likely be attributed to the reduced availability of food due to the loss in food production among affected households, which was uncovered by our method and Bonferroni correction at $\Gamma_{\c}=1$, along with the growth in food price index, which was detected by the Sens-Val and Na\"ive methods until $\Gamma_{\c}=1.5$. Likewise, all approaches discover a significant increase in budget share used to buy pulses, such as beans, lentils, and peas, which was also alluded to in the report. Sens-Val and Bonferroni correction also identify a decrease in rice consumption as an effect of the flooding.

Moreover, the Sens-Val procedure uncovers the discrepancy in the proportion of households that receive their washing water from tubewells until $\Gamma_{\c}=1.5$ and the proportion of households that receive their cooking water from tubewells until $\Gamma_{\c}=2.5$. However, the Na\"ive method and full-sample Bonferroni correction fail to detect these effects at the same levels of allowance for potential biases, further demonstrating the robustness and power of our new technique. Since tubewells are the only source of clean water accessible to the vast majority of rural inhabitants in Bangladesh (\cite{del20011998}), residents were thus more likely to fall ill due to poor water quality and sanitation, which has been heavily documented following floods and other natural disasters (\cite{centers2000morbidity,du2010health}). Indeed, both the Sens-Val and Na\"ive methods find that households exposed to the floods experienced a larger number of days with illness compared to their matched counterparts until $\Gamma_{\c}=1.5$. Although we were able to identify this augmented illness intensity among exposed households, no significant evidence of long-term nutritional detriments was discovered. This result is partially in-line with a later work by \cite{del2005treading} which found that many of these metrics had recovered one year after the floods.

Our Sens-Val approach consistently selects more outcomes in the planning sample to test compared to the Na\"ive procedure. Far from being a bug, this was a primary reason for our introduction of a more nuanced methodology. The Na\"ive method often suffers from lesser power due to filtering out promising outcomes in the planning stage, despite their potential to prove significant in the subsequent part of the pipeline. On the other hand, our method rejects at least as many outcomes as those rejected by Bonferroni correction for each value of $\Gamma_{\c}.$ Our approach also demonstrates enhanced power at large values of $\Gamma_{\c}$. In particular, Bonferroni correction fails to recover any outcomes for $\Gamma_{\c} \geq 16$ and the Na\"ive method fails to do so for $\Gamma_{\c} \geq 20$. Yet Sens-Val still rejects ft.water, repair.cost, and days.water at $\Gamma_{\c} = 38$. These outcomes are all quite intuitively related to our treatment exposure and it is thus encouraging to see our Sens-Val method identifying their significance even when accounting for very large potential biases.

\label{practical}

\section{Discussion} \label{discussion}

In this work, we introduce a method to determine how many and which outcomes to select from a pilot sample based on the principle of prioritizing outcomes that are more resilient to potential hidden biases in an observational study. We consider both the sensitivity value and its corresponding variability for each outcome in the planning sample to construct predictive intervals on the analysis sample and render a robust selection of hypotheses that have a reasonable chance of being rejected. One possible direction to extend Method \ref{bootstrap} is inspired by the results in Section \ref{edgeworth-section}. Since the Edgeworth expansion delivers a confidence interval more accurate than the classical central limit theorem, one could obtain prediction intervals that are accurate up to $O(I^{-1})$. Indeed, one can easily verify that $r^{1/2}V^l_{\p} - (1-r)^{1/2}V^l_{\a}$ equals the statistic that we use in Method \ref{bootstrap}, up to $O(I^{-1})$, where the subscripts denote the counterparts of $V^l$ from the planning and analysis samples. By virtue of the independence of these samples, the Edgeworth expansion of this linear combination can be obtained, albeit by tedious calculation, to render a form similar to that in Theorem \ref{edgeworth}. The approximate quantiles of that distribution can be obtained by estimating the coefficients of the $I^{-1/2}$ term, which in turn can be estimated at $I^{1/2}$ rate via bootstrapping from the planning sample, yielding a $O(I^{-1})$ approximation of the quantiles and a corresponding coverage probability of $(1-\alpha_{\cv}) + O(I^{-1})$, improving upon Corollary \ref{ourcor1}. %

We present our method as a means for securing FWER control, thereby rendering the Bonferroni procedure a clear baseline for comparison; however our strategy is highly flexible and can be adapted easily to alternative criteria. For instance, the false discovery rate (FDR) is another metric for which control is often sought when conducting multiple comparisons. The FDR can be controlled during the variable selection process by techniques such as the Benjamini–Hochberg procedure (\cite{benjamini1995controlling}) and model-X knockoffs (\cite{candes2018panning}), the latter of which helped to inspire the title of this article. 

{In practice, the value of $\Gamma_{\c}$ is typically pre-specified based on the knowledge and judgment of the researcher or subject matter expert, but doing so may be difficult without relevant prior literature or external calibration. As an alternative, one may explore how the Sens-Val filter behaves across a range of candidate values for $\Gamma_{\c}$ on the planning sample. By plotting the number of outcomes retained as a function of $\Gamma_{\c}$, the resulting scree plot reveals how quickly the filtering tightens as sensitivity is increased, and a reasonable candidate of $\Gamma_{\c}$ can be chosen at the threshold where the curve begins to level off.}

A possible future direction of this work could be to adapt the proposed method to the framework of cross-screening (\cite{zhao2018cross}) or data turnover (\cite{bekerman2024protocol}) to {combine evidence from both the planning and analysis samples}. Cross-screening randomly splits in half some data, uses each half to plan and execute a study on the other half, and then reports the more favorable results, adjusting for multiple comparisons using the Bonferroni correction. {Data turnover builds on the cross-screening framework to allow for certain data-informed insights to be incorporated into the analysis plan. Likewise, one may seek to adapt classical sequential analysis methods for two-stage designs (e.g., \cite{bauer1994evaluation}) to integrate with our hypothesis screening procedure and make fuller use of both samples for inference.
    While the primary focus of this paper is to describe our procedure using traditional splitting-based inferences, %
    it could be an interesting future direction to explore whether it is advantageous to employ our design filter together with these approaches}, especially when there are many pre-specified hypotheses of interest.

\section*{Acknowledgement}
The authors thank Ting Ye and the anonymous referees for their helpful ideas, insightful comments, and constructive suggestions. Bekerman was partially supported by the National Science Foundation Graduate Research Fellowship. Small was partially supported by the National Institutes of Health.

\section*{Reproducibility}
\label{SM}
Bangladesh flooding data is publicly available (\texttt{https://doi.org/10.7910/DVN/UJA9N5}) and an \texttt{R} package for implementing our method, reproducing numerical experiments, and conducting data analysis is available on \texttt{GitHub} (\texttt{https://github.com/WillBekerman/planning-for-gold}).

\bibliographystyle{biometrika}
\bibliography{references.bib}

\newpage

\appendix
\section{Technical Proofs}

\vspace{.75em}
\subsection{Proof of Theorem \ref{ourmain1}}
We suppress the superscript $l$ in the proof for ease of notation. Using Equation \ref{Qing-main}, after sufficient enriching of the underlying probability space, we have
\begin{align*}
    \kappa_{\alpha_{\p}} - \mu_F &= -\dfrac1{\sqrt{I_{\p}}}{\sigma_q\Phi^{-1}(1-\alpha_{\p})\sqrt{\mu_F(1-\mu_F)}} + \dfrac{Z_1\sigma_F}{\sqrt {I_{\p}}} + o_p\left(\dfrac{1}{\sqrt{I_{\p}}}\right)\\
    \kappa_{\alpha_l} - \mu_F &= -\dfrac1{\sqrt{I_{\a}}}{\sigma_q\Phi^{-1}(1-\alpha_l)\sqrt{\mu_F(1-\mu_F)}} + \dfrac{Z_2\sigma_F}{\sqrt{I_{\a}}} + o_p\left(\dfrac{1}{\sqrt{I_{\a}}}\right),
\end{align*}
where $Z_1$ and $Z_2$ are independent standard Normals. Subtracting one from the other and noting that $I_{\p} =rI, I_{\a} = (1-r)I$, we obtain 
\begin{align*}
    \kappa_{\alpha_l} - \kappa_{\alpha_{\p}} &= \dfrac{\sqrt{\mu_F(1-\mu_F)}\sigma_q}{\sqrt I}\left[\dfrac{\Phi^{-1}(1-\alpha_{\p})}{\sqrt r} - \dfrac{\Phi^{-1}(1-\alpha_l)}{\sqrt {1-r}}\right] - \dfrac{\sigma_F}{\sqrt I}\left(\dfrac{Z_1}{\sqrt r} - \dfrac{Z_2}{\sqrt{1-r}} \right)\\ &+ o_p\left(\dfrac 1{\sqrt I} \right)
\end{align*}
Finally, noting that $\kappa_{\alpha_{\p}} = \mu_F + o_p(1)$, we find that 
\begin{align*}\kappa_{\alpha_l} &= \kappa_{\alpha_{\p}} + \dfrac{\sqrt{\kappa_{\alpha_{\p}}(1-\kappa_{\alpha_{\p}})}\sigma_q}{\sqrt I}\left[\dfrac{\Phi^{-1}(1-\alpha_{\p})}{\sqrt r} - \dfrac{\Phi^{-1}(1-\alpha_l)}{\sqrt {1-r}}\right] - \dfrac{\sigma_F}{\sqrt{I}}\left(\dfrac{Z_1}{\sqrt r} - \dfrac{Z_2}{\sqrt{1-r}}\right) \\ & \hspace{2em}+ o_p\left(\dfrac 1{\sqrt I} \right)
\end{align*}
Since $Z_1/\sqrt r - Z_2/\sqrt{1-r}$ has a $\mathcal{N}(0, 1/r(1-r))$ distribution, our proof is complete. \hfill $\qed$

\subsection{Proof of Proposition \ref{power}}
{We suppress the superscript $l$ for clarity. The Na\"ive method selects outcome $l$ when $\bar p_{\Gamma_{\c}}\le \alpha$, or equivalently, when $\kappa_{\alpha_{\p}} \ge \kappa_{\c}.$ On the other hand, the Sens-Val method selects an outcome $l$ iff 
$$\kappa_{\alpha_{\p}} \ge \kappa_{\c} - \dfrac{\sqrt{\kappa_{\alpha_{\p}}(1-\kappa_{\alpha_{\p}})}\sigma_q}{\sqrt{I}}\left(\dfrac{\Phi^{-1}(1-\alpha_{\p})}{\sqrt{r}} - \dfrac{\Phi^{-1}(1-\alpha_l)}{\sqrt{1-r}}\right) - \dfrac{\Phi^{-1}(1-\alpha_{\cv})\hat\sigma_F}{\sqrt{r(1-r)I}}.$$
Thus to ensure that if outcome $l$ is selected by the Na\"ive method it is also selected by Sens-Val, it is enough to ensure that 
$$\kappa_{\c} - \dfrac{\sqrt{\kappa_{\alpha_{\p}}(1-\kappa_{\alpha_{\p}})}\sigma_q}{\sqrt{I}}\left(\dfrac{\Phi^{-1}(1-\alpha_{\p})}{\sqrt{r}} - \dfrac{\Phi^{-1}(1-\alpha_l)}{\sqrt{1-r}}\right) - \dfrac{\Phi^{-1}(1-\alpha_{\cv})\hat\sigma_F}{\sqrt{r(1-r)I}} \le \kappa_{\c}$$ or equivalently,
$$\dfrac{\sqrt{\kappa_{\alpha_{\p}}(1-\kappa_{\alpha_{\p}})}\sigma_q}{\sqrt{I}}\left(\dfrac{\Phi^{-1}(1-\alpha_{\p})}{\sqrt{r}} - \dfrac{\Phi^{-1}(1-\alpha_l)}{\sqrt{1-r}}\right) + \dfrac{\Phi^{-1}(1-\alpha_{\cv})\hat\sigma_F}{\sqrt{r(1-r)I}}\ge 0.$$
To ensure this, it is enough to ensure that
$$\dfrac{\Phi^{-1}(1-\alpha_{\p})}{\sqrt{r}} \ge \dfrac{\Phi^{-1}(1-\alpha_l)}{\sqrt{1-r}} \ \forall l$$
or equivalently $$r\le \inf_{l\in [L]}\dfrac{[\Phi^{-1}(1-\alpha_{\p})]^2}{[\Phi^{-1}(1-\alpha_{\p})]^2 + [\Phi^{-1}(1-\alpha_{l})]^2},$$
which can always be ensured by choosing $r$ appropriately.}

{{Remark}: If $\alpha_l$ is chosen to be $\alpha/L$ as a hyperparameter for the Sens-Val method, then the above condition boils down to $$L\le \dfrac{\alpha}{1-\Phi\left(\sqrt{\dfrac{1-r}{r}}\Phi^{-1}(1-\alpha_{\p})\right)}.$$ For $\alpha_{\p} = \alpha =0.05$, the right-hand side evaluates to $\approx 22.8$ when $r = 0.25$, $\approx 99.7$ at $r = 0.2$ and $\ge 1100$ at $r=0.15$, and hence one can suitably choose the hyperparameters based on the number of outcomes.}

\subsection{Proof of Proposition \ref{vble-expand}}

We state and prove the extended version of Proposition \ref{vble-expand}. Suppose the conditions of Equation \ref{clt} and Equation \ref{alt-clt} hold. Additionally, assume $c^l_{q,I} = {\frac 1n\sum_{i=1}^I (q^l_i)^3}/{\left(\frac 1n\sum_{i=1}^n q^l_i\right)^3}$ is such that $c_q^l = \lim_{I\to\infty} c^l_{q,I} <\infty$. Define $T_{\text{alt}}^l = {\sqrt{I}(T^l - \mu_F^l)}/{\sigma_F^l}$. Then,
\begin{itemize}
    \item[(i)] $V^l := \dfrac{\sqrt I(\kappa_\alpha^l - \mu_F^l)}{\sigma_F^l} + \dfrac{z_\alpha\sigma^l_{q,I}}{\sigma_F^l}\sqrt{\kappa_\alpha^l(1-\kappa_\alpha^l)} = T^l_{\text{alt}} - (2\mu_F^l-1)\dfrac{c_1}{\sqrt{I}}+ O_p\left(\dfrac 1I\right)$
    \item[(ii)] $W^l:= \dfrac{\sqrt{I}(\kappa_\alpha^l - \mu_F^l)}{\sigma_F^l}$ admits the expression \begin{align*} W^l &= T^l_{\text{alt}}\left[1+ \dfrac{\sigma^l_{q,I}z_\alpha(2\mu_F^l-1)}{2\sqrt{I}\sqrt{\mu_F^l(1-\mu_F^l)}}\right] - \dfrac{\sigma^l_{q,I}z_\alpha}{\sigma_F^l}\sqrt{\mu_F^l(1-\mu_F^l)} - \dfrac{2\mu_F^l-1}{\sqrt I}\left[c_1 + \dfrac{(\sigma_{q,I}^l)^2}{\sigma_F^l}z_\alpha^2\right] \\ &\hspace{30em}+ O_p\left(\dfrac 1I\right) 
    \vspace{-1em} \end{align*}
\end{itemize}
where $c_1 = c_{q,I}^lg(z_\alpha)/\sigma_F^l$ and $g(x) = (x^2-1)/6$.
We suppress the subscript $l$ as understood from context and start by proving Proposition \ref{vble-expand} (i). Referring to the distribution of $T_\kappa := \sqrt{I}(\bar T_\Gamma - \kappa)/\sqrt{\kappa(1-\kappa)\sigma_{q,I}}$ as $G_{I,\kappa}$, where $\kappa=\Gamma/(1+\Gamma)$, if $c_{q,I} = {\frac 1n\sum_{i=1}^I q_i^3}/{\left(\frac 1n\sum_{i=1}^n q_i\right)^3}$ is such that $c_q = \lim_{I\to\infty} c_{q,I} <\infty$, the quantiles of $G_{I,\kappa}$ admit a Cornish-Fisher expansion (\cite{DasGupta2008}) of the form $$G_{I,\kappa}^{-1}(t) = \Phi^{-1}(t) + \dfrac{\kappa_3}{\kappa_2^{3/2}}g(\Phi^{-1}(t)) + O\left(\dfrac 1I\right)$$ where $\kappa_r$ is the $r$-th cumulant of $T_\kappa$, $r\ge 1$, and $g(x) = \frac 16(x^2-1)$. Now, by the scaling chosen for $T_\kappa$, we see that $\kappa_2 = \V(T_\kappa) = 1$ and a quick calculation reveals $$\kappa_3 = \E[T_\kappa^3] = \dfrac 1{\sqrt{I}} \dfrac{2\kappa-1}{\sqrt{\kappa(1-\kappa)}\sigma_{q,I}}c_{q,I} = O(1) \ \ (\text{by assumption})$$
Now the transformed sensitivity value $\kappa_{\alpha}$ satisfies 
\begin{align}
    \dfrac{\sqrt I(T-\kappa_{\alpha})}{\sqrt{\kappa_{\alpha}(1-\kappa_{\alpha})}\sigma_{q,I}} &= G_{I,\kappa_{\alpha}}^{-1}(1-\alpha) = z_{\alpha} + \dfrac{1}{\sqrt I}\dfrac{2\kappa_{\alpha}-1}{\sqrt{\kappa_{\alpha}(1-\kappa_{\alpha})}\sigma_{q,I}}c_{q,I}g(z_{\alpha}) + O_p\left(\frac 1I\right) \nonumber \\
   \implies \dfrac{\sqrt I(T-\mu_F)}{\sigma_F}&= \dfrac{\sqrt{I}(\kappa_{\alpha}-\mu_F)}{\sigma_F} + z_{\alpha}\dfrac{\sigma_{q,I}}{\sigma_F}\sqrt{\kappa_{\alpha}(1-\kappa_{\alpha})} + \dfrac{2\kappa_{\alpha} - 1}{\sqrt I}\dfrac{c_{q,I}g(z_{\alpha})}{\sigma_F} + O_p\left(\frac 1I\right) \label{eq1}
\end{align}
Finally, noting that $\kappa_{\alpha} = \mu_F + O_p(1/\sqrt{I})$ we thus obtain from Equation \ref{eq1}
$$V = T_{\text{alt}} - (2\mu_F-1)\dfrac{c_1}{\sqrt I} + O_p\left(\dfrac 1I\right)$$ as claimed in Proposition \ref{vble-expand} (i).

Next, we prove Proposition \ref{vble-expand} (ii) and continue from Equation \ref{eq1}. Using the variables defined above, we have 
\begin{align*}
    T_{\text{alt}} &= W + \dfrac{\sigma_{q,I}z_{\alpha}}{\sigma_F}\sqrt{\left(\mu_F+\dfrac{W\sigma_F}{\sqrt I}\right)\left(1-\mu_F-\dfrac{W\sigma_F}{\sqrt I}\right)} + \left(\dfrac{2W\sigma_F}{\sqrt I} + 2\mu_F-1\right)\dfrac{c_1}{\sqrt I} + O_p\left(\dfrac 1I\right)\\
    &= W + \dfrac{\sigma_{q,I}z_{\alpha}}{\sigma_F}\sqrt{\mu_F(1-\mu_F) - \dfrac{W\sigma_F}{\sqrt I}(2\mu_F-1)} + (2\mu_F-1)\dfrac{c_1}{\sqrt I} + O_p\left(\dfrac 1I\right)
\end{align*}
Next, using $\sqrt{a+x}-\sqrt a = \frac{x}{2\sqrt a} + O(x^2)$ we obtain
\begin{align*} T_{\text{alt}} = W\left[1- \dfrac{\sigma_{q,I}z_{\alpha}(2\mu_F-1)}{2\sqrt I\sqrt{\mu_F(1-\mu_F)}}\right] + \dfrac{\sigma_{q,I}z_{\alpha}}{\sigma_F}\sqrt{\mu_F(1-\mu_F)}  + (2\mu_F-1)\dfrac {c_1}{\sqrt I} +O_p\left(\frac 1I\right). \end{align*}
Continuing with the trail of equalities
\begin{align}
    W&=\left[1- \dfrac{\sigma_{q,I}z_{\alpha}(2\mu_F-1)}{2\sqrt I\sqrt{\mu_F(1-\mu_F)}}\right]^{-1}\left[ T_{\text{alt}} - \dfrac{\sigma_{q,I}z_{\alpha}}{\sigma_F}\sqrt{\mu_F(1-\mu_F)} - (2\mu_F-1)\dfrac{c_1}{\sqrt I}\right] + O_p\left(\dfrac 1I\right) \nonumber \\
    &= \left[1+ \dfrac{\sigma_{q,I}z_{\alpha}(2\mu_F-1)}{2\sqrt I\sqrt{\mu_F(1-\mu_F)}} + O_p\left(\frac 1I\right)\right]\left[ T_{\text{alt}} - \dfrac{\sigma_{q,I}z_{\alpha}}{\sigma_F}\sqrt{\mu_F(1-\mu_F)} - (2\mu_F-1)\dfrac{c_1}{\sqrt I}\right] \nonumber \\ &+ O_p\left(\dfrac 1I\right)\nonumber \\
    \therefore W &= T_{\text{alt}}\left[1+ \dfrac{\sigma_{q,I}z_{\alpha}(2\mu_F-1)}{2\sqrt I\sqrt{\mu_F(1-\mu_F)}}\right] - \dfrac{\sigma_{q,I}z_{\alpha}}{\sigma_F}\sqrt{\mu_F(1-\mu_F)} - \dfrac{2\mu_F-1}{\sqrt I}\left[c_1  + \dfrac{\sigma_{q,I}^2z_{\alpha}^2}{2\sigma_F}\right]\\ &+ O_p\left(\frac 1I\right) \nonumber ,
\end{align}
concluding the proof. Moreover, it can be seen from Proposition \ref{vble-expand}(ii) that if $I\to\infty$, one recovers the asymptotic result given as \ref{Qing-main}. \hfill $\qed$

\subsection{Proof of Theorem \ref{edgeworth}}
We state and prove the extended version of Theorem \ref{edgeworth}. Suppose the conditions of Proposition \ref{vble-expand} and Equation \ref{edge-ass} hold. Then, 
    \begin{itemize}
        \item[(i)] $\P(V^l\le v)= \Phi(v) + \dfrac{1}{\sqrt I}\phi(v)\left[(2\mu_F^l-1)c_1 + b^l(v,F)\right] + O\left(\dfrac 1I\right)$
        \item[(ii)] 
            $\begin{aligned}[t]
            \P(W^l\le w) &= \Phi\left(w + \frac{\sigma^l_{q,I}z_\alpha}{\sigma^l_F}\sqrt{\mu_F^l(1-\mu^l_F)}\right) \\ & + \dfrac{1}{\sqrt I}\phi\left(w + \frac{\sigma_{q,I}^lz_\alpha}{\sigma_F^l}\sqrt{\mu_F^l(1-\mu_F^l)}\right)\left[(2\mu_F^l-1)\left(c_1 + \frac{(\sigma^l)^2_{q,I}z^2_\alpha}{2\sigma_F^l}\right)  \right.\\ &\hspace{1em}\left. -\left(w + \frac{\sigma^l_{q,I}z_\alpha}{\sigma^l_F}\sqrt{\mu^l_F(1-\mu^l_F)}\right)\left(\frac{(2\mu^l_F - 1)\sigma_q^lz_\alpha}{2\sqrt{\mu_F^l(1-\mu_F^l)}}\right) + b^l\left(w + \dfrac{\sigma^l_{q,I}z_\alpha}{\sigma^l_F}, F\right)\right] \\ &+ O\left(\dfrac 1I\right)
            \end{aligned}$ %
    \end{itemize}
Analogous expressions hold for the bootstrapped version with $\P$ replaced by $\P_*$, $\mu_F$ and $F$ replaced by $\kappa_\alpha$ and empirical distribution $\hat F$, respectively. We provide a proof of Theorem \ref{edgeworth} (i) as (ii) follows similarly. We suppress the superscript $l$ and prove the version for $\P$, with the version for $\P_*$ being analogous. From Proposition \ref{vble-expand} and Equation \ref{edge-ass}, %
we obtain 
\begin{align*}
    \P(V\le v) &= \P\left(T_{\text{alt}}\le v + (2\mu_F-1)\dfrac{c_1}{\sqrt I} + O_p\left(\dfrac1I\right)\right)\\
    &= \Phi\left(v + (2\mu_F-1)\dfrac{c_1}{\sqrt I} + O\left(\dfrac1I\right)\right) \\ &\hspace{2em}+ \dfrac{1}{\sqrt I}\phi\left(v + (2\mu_F-1)\dfrac{c_1}{\sqrt I} + O\left(\dfrac1I\right)\right)b\left(v + (2\mu_F-1)\dfrac{c_1}{\sqrt I} + O\left(\dfrac1I\right), F\right) + O\left(\dfrac 1I\right)
\end{align*}
{The move from $O_p(\cdot)$ to $O(\cdot)$ follows from a concentration inequality of the form $$\P\left(\left|\sqrt{\kappa_\alpha(1-\kappa_\alpha)} - \sqrt{\mu_F(1-\mu_F)}\right|>\delta\right)\le C\exp(-cI\delta^2),$$ which is a consequence of a similar concentration inequality for $T$ obtained from the assumed Edgeworth expansion of $T$.}

Finally, using $f(x+O(1/\sqrt{I})) = f(x) + O(1/\sqrt I)$, for any differentiable function $f$ and $\Phi(a+x/\sqrt{I}) = \Phi(a) + \phi(a)x/\sqrt{I} + O(1/I)$ we deduce 
$$\P(V\le v) = \Phi(v) +\dfrac{1}{\sqrt I}[(2\mu_F - 1)c_1+b(v,F)]\phi(v) + O\left(\dfrac 1I\right),$$ completing the proof.\hfill $\qed$

{{Remark}: The Edgeworth expansion for $V$ has an additional term of $(2\mu_F-1)$ for the $1/\sqrt{I}$ term, which is a consequence of using the Cornish-Fisher expansion (and a non-asymptotic null distribution). If one instead uses the asymptotic normal distribution to obtain the null, the equation to be satisfied would be $$\sqrt{I}(T-\kappa_\alpha) = z_\alpha \sqrt{\kappa_\alpha(1-\kappa_\alpha)}\sigma_{q,I},$$ and the additional terms would not appear in the expansion of $V$.}

\subsection{Proof of Proposition \ref{le-cam}}
We suppress the superscript $l$ throughout the proof.
$$T_I = \dfrac{\sum_{i=1}^I \psi(IG_I(|Y_i|/(I+1))\text{sgn}(Y_i)}{\sum_{i=1}^I \psi(IG_I(|Y_i|)/(I+1))} $$ where $G_I$ is the empirical CDF of $I|Y_i|/(I+1)$. We have
\begin{align*} \sqrt{I}T_I &= \dfrac{\frac{1}{\sqrt I}\sum_{i=1}^I \psi(IG_I(|Y_i|/(I+1))\text{sgn}(Y_i)}{\frac 1I \sum_{i=1}^I \psi(IG_I(|Y_i|/(I+1))} = \dfrac{1}{{\sqrt I\|\psi\|_1}}\sum_{i=1}^I \psi(IG_I(|Y_i|/(I+1))\text{sgn}(Y_i) + o_p(1) \\ &=
\dfrac{1}{\sqrt{I}\|\psi\|_1}\sum_{i=1}^I \psi(G(|Y_i|)\text{sgn}(Y_i) + \dfrac{1}{\sqrt{I}\|\psi\|_1}\sum_{i=1}^I \left[\psi\left(\dfrac{I}{I+1}G_I(|Y_i|)\right) - \psi(G(|Y_i|))\right]\text{sgn}(Y_i) \\ & \hspace{4em}+ o_p(1)
\end{align*}
where $G$ is the CDF of $|Y_i|$. Call the second term $D_I/\|\psi\|_1$. Applying Hoeffding's inequality, we have
\begin{align}
    \P(|D_I|>\varepsilon) &= \E\left[\P\left(\left.\dfrac{1}{\sqrt I}\sum_{i=1}^I \left[\psi\left(\dfrac{I}{I+1}G_I(|Y_i|)\right) - \psi(G(Y_i))\right]\text{sgn}(Y_i)>\varepsilon \right| |Y_1|,\cdots,|Y_I|\right)\right] \nonumber \\
    &\le \E\left[2\exp\left(-\dfrac{I\varepsilon^2}{2\sum_{i=1}^I \left[\psi\left(\dfrac{I}{I+1}G_I(|Y_i|)\right) - \psi(G(|Y_i|))\right]^2}\right)\right] \nonumber \\
    &\le \E\left[2\exp\left(-\dfrac{I\varepsilon^2}{2B^2\sum_{i=1}^I \left[\dfrac{I}{I+1}G_I(|Y_i|) -G(|Y_i|)\right]^2}\right)\right] \label{hoeff}
\end{align}
where $B = \sup_{x\in (0,1)} \psi'(x)<\infty$. Now, we observe 
\begin{align*}\dfrac{1}{I}\sum_{i=1}^I \left[\dfrac{I}{I+1}G_I(|Y_i|) - G(|Y_i|)\right]^2&\le \dfrac 1I\sum_{i=1}^I 2\left[\dfrac{I^2}{(I+1)^2}(G_I(|Y_i|) - G(|Y_i|))^2 + \dfrac{1}{(I+1)^2}G(|Y_i|)^2\right]\\
&\le \dfrac{2I^2}{(I+1)^2}\|G_I - G\|_\infty + \dfrac{2}{(I+1)^2}\dfrac{1}{I}\sum_{i=1}^I G(|Y_i|)^2\end{align*}
which converges to 0 by Glivenko-Cantelli theorem and the weak law of large numbers since $\E[G(|Y|)^2]<\infty$. Noting that $e^{-x}\le 1$ for all $x\ge 0$, by bounded convergence theorem on Equation \ref{hoeff}, we obtain $D_I = o_p(1)$.

Thus, we have 
$$\dfrac{\sqrt I}{\sigma(\theta_0)}(T_I - \mu(\theta_0)) = \dfrac{1}{\sqrt I}\dfrac{\sum_{i=1}^I U_i(\text{sgn}(Y_i) - \mu(\theta_0))}{\sigma(\theta_0)\|\psi\|_1} +o_p(1),$$ where $U_i = \psi(G(|Y_i|))$ and $G$ is the CDF of $|Y_i|$, thereby allowing $T_I$ to have an asymptotic linear expansion at $\theta_0$. Also, as the location family has a continuously differentiable positive density, the Fisher information derived at $\theta_0$ is $I = \int \dfrac{f'(x-\theta_0)^2}{f(x-\theta_0)}dx = \int \frac{f'(x)^2}{f(x)}\,dx$ is constant and thus continuous at $\theta_0$. Hence, the family is quadratic mean differentiable (\cite{van2000asymptotic}, Theorem 7.2), leading to 
$\sqrt{I}(T_I - \mu(\theta_0))/\sigma(\theta_0)$ and the log-likelihood ratio at $\theta_I$ vs $\theta_0$ satisfying 
$$\begin{bmatrix} \sqrt{I}(T_I - \mu(\theta_0))/\sigma(\theta_0) \\  \log\dfrac{\prod_{i=1}^I f_{\theta_0+ h/\sqrt I}(Y_i)}{\prod_{i=1}^I f_{\theta_0}(Y_i)}\end{bmatrix} = \dfrac{1}{\sqrt I}\begin{bmatrix}\displaystyle\sum_{i=1}^I\dfrac{ U_i(\text{sgn}(Y_i) - \mu(\theta_0))}{\sigma(\theta_0)\|\psi\|_1}\\ h\displaystyle\sum_{i=1}^I\dfrac{ f'_{\theta_0}(Y_i)}{f_{\theta_0}(Y_i)}\end{bmatrix} + \begin{bmatrix} 0 \\ -h^2I/2\end{bmatrix} + \begin{bmatrix} o_p(1) \\ o_p(1)\end{bmatrix}$$
and thus converging to a bivariate normal distribution. %
By Le-Cam's third lemma (\cite{van2000asymptotic}, Section 6.2), our proof is complete.\hfill $\qed$

\subsection{Proof of Theorem \ref{localasympot}}

An outcome $l$ is selected by Method \ref{bootstrap} if its corresponding
\begin{align*}
    & \kappa_{\alpha_{\p}} + \dfrac{\sqrt{\kappa_{\alpha_{\p}}(1-\kappa_{\alpha_{\p}})}\sigma_qz_{\alpha_{\p}}}{\sqrt{Ir}} - \dfrac{\sqrt{\kappa_{\alpha_{\p}}(1-\kappa_{\alpha_{\p}})}\sigma_qz_{\alpha_l}}{\sqrt{I(1-r)}} > \kappa_{\c} - \dfrac{\hat\sigma(\hat\theta_I)z_{\alpha_{\cv}}}{\sqrt{Ir(1-r)}} \\
    \iff & \sqrt{Ir}\left(\kappa_{\alpha_{\p}} + \dfrac{\sqrt{\kappa_{\alpha_{\p}}(1-\kappa_{\alpha_{\p}})}\sigma_qz_{\alpha_{\p}}}{\sqrt{Ir}} - \dfrac{\sqrt{\kappa_{\alpha_{\p}}(1-\kappa_{\alpha_{\p}})}\sigma_qz_{\alpha_l}}{\sqrt{I(1-r)}}\right) > \\ &\hspace{2em}\sqrt{Ir}\kappa_{\c} - \dfrac{\hat\sigma(\hat\theta_I)z_{\alpha_{\cv}}}{\sqrt{1-r}}\\
    \iff &\sqrt{I_{\p}}\dfrac{(\kappa_{\alpha_{\p}}- \mu(\theta_I))}{\sigma(\theta_I)} + \sqrt{\kappa_{\alpha_{\p}}(1-\kappa_{\alpha_{\p}})}\dfrac{\sigma_q}{\sigma(\theta_I)}z_{\alpha_{\p}} - \dfrac{\sqrt{\kappa_{\alpha_{\p}}(1-\kappa_{\alpha_{\p}})}\sqrt{r}}{\sqrt{(1-r)}}\dfrac{\sigma_q}{\sigma(\theta_{I})}z_{\alpha_l} > \\
    &\hspace{2em} \sqrt{I_{\p}}\dfrac{(\kappa_c - \mu(\theta_I))}{\sigma(\theta_I)} -\dfrac{\hat\sigma(\hat\theta_I)}{\sigma(\theta_I)}\dfrac{z_{\alpha_{\cv}}}{\sqrt{1-r}} 
\end{align*}
the probability of which, under $\theta_I$ and as $I\to\infty$, is given by
$$\pi_I(\theta_I)\to 1-\Phi\left(-\dfrac{z_{\alpha_{\cv}}}{\sqrt{1-r}} - h\dfrac{\mu'(\theta_0)}{\sigma(\theta_0)} + z_{\alpha_{l}}\dfrac{\sqrt{\kappa(1-\kappa)}\sigma_q}{\sigma(\theta_0)}\dfrac{\sqrt r}{\sqrt{1-r}}\right)$$
as intended.\hfill $\qed$

\subsection{Proof of Theorem \ref{sens-multi}}

We suppress the $l$-superscript throughout the proof and drop mention of $\bm Z$ and $\bm R^l$. %
If the sensitivity analysis is computed via Equation \ref{multi-ass}, then $\Gamma^*_{\alpha}$ satisfies Equation \ref{rep1}, i.e.,
\begin{align}
    T - a_{\Gamma^*_{\alpha}} &= \Phi^{-1}(1-\alpha)\dfrac{b_{\Gamma^*_{\alpha}}}{\sqrt I} + o_p\left(\dfrac{1}{\sqrt I}\right). \label{rep2}
\end{align}
Let $f$ be the mapping $f:\Gamma\mapsto a_{\Gamma}$, which is continuously differentiable at a neighborhood of $\tilde\Gamma$ with a non-zero derivative at $\tilde\Gamma$ by assumption. By inverse function theorem (\cite{rudin1976principles}, Theorem 9.24), $f$ is invertible at neighborhood of $\tilde\Gamma$ and the inverse also admits a derivative. 
\newline 
Since $T = \mu_{F,g} + o_p(1)$, rejection of the null hypothesis occurs w.p. 1 as $I\to\infty$ if $\Gamma$ is such that $a_\Gamma < \mu_{F,g}$. On the other hand, rejection of the null hypothesis occurs w.p. 0 as $I\to\infty$ if $\Gamma$ is such that $a_\Gamma>\mu_{F,g}$. Thus, by definition, $a_{\tilde\Gamma} = \mu_{F,g}$ implies $\tilde\Gamma = f^{-1}(\mu_{F,g})$, where the inverse is well-defined in a neighborhood of $\mu_{F,g}$ by the inverse function theorem.
\newline 
Since $T\overset{p}{\to}\mu_{F,g}$, we obtain $a_{\Gamma^*_{\alpha}}\overset{p}{\to} \mu_{F,g}$ from Equation \ref{rep2} and deduce $\Gamma_{\alpha}^*\overset{p}{\to} \tilde\Gamma^l$ by continuous mapping theorem. Using Equation \ref{rep1} and Assumption \ref{alt-clt-multi}, we can obtain 
\begin{align*}
    \dfrac{\sqrt I(a_{n,\Gamma^*_{\alpha}} - \mu_{F,g})}{\sigma_{F,g}} &= \dfrac{\sqrt{I}(T-\mu_{F,g})}{\sigma_{F,g}} - \Phi^{-1}(1-\alpha)\dfrac{b_{\Gamma_{\alpha}^*}}{\sigma_{F,g}} + o_p(1)
\end{align*}
Replacing $b_{\Gamma_{\alpha}^*}$ by $b_{\tilde\Gamma}$ and using continuous mapping theorem, we have
$$\dfrac{\sqrt{I}(a_{\Gamma_{\alpha}^*} - \mu_{F,g})}{\sigma_{F,g}} \overset{d}{\to}\mathcal{N}\left(-\Phi^{-1}(1-\alpha)\dfrac{b_{\tilde\Gamma}}{\sigma_{F,g}}, 1\right)$$
Thus, by the inverse-function theorem and delta method using the function $f^{-1}$ in a neighborhood of $\mu_{F,g}$ we conclude 
$$\sqrt I(\Gamma^*_{\alpha} - \tilde\Gamma)\overset{d}{\to}\mathcal{N}\left(-\Phi^{-1}(1-\alpha)b_{\tilde\Gamma} (f^{-1})'(\mu_{F,g}),\sigma^2_{F,g}((f^{-1})'(\mu_{F,g}))^2\right)$$ \hfill $\qed$

\section{{Full Matching}} \label{sec:multi}
{We extend our results to full matching, a general approach which {allows for each matched set to consist of multiple treated units or multiple control units. Following \cite{rosenbaum1991characterization}, we adopt the convention that a full matching consists of matched sets such that each set has either one treated unit and at least one control unit, or one control unit and at least one treated unit. }
    This facilitates a fuller use of the data, as full matching incorporates all available subjects, while typically achieving reasonable covariate balance across treatment and comparison groups. Notably, full matching accommodates regions of the covariate space where treated units are sparse relative to controls (or vice versa), thereby retaining subjects that other matching schemes might otherwise exclude.} 
    
    {In this section, we introduce Theorems \ref{sens-multi} and \ref{ourmain-multi}, and Method \ref{bootstrap-multi}, that generalize Sens-Val to full matching. While delineated in the most general case, we note that these results apply also to fixed-ratio matching (including matching with multiple controls) and variable ratio-matching as special cases. 

{Formally, we consider} the setting with $n_i\ge 2$ units per matched set, with the $i$-th matched set either containing one treated unit and $n_i - 1$ control units or one control and $n_i-1$ treated units}, and we may be interested in the treated-minus-control difference $Y_{ij}^l$ for the $l$th outcome. For instance, if $n_1 = 3$ {with one treated unit and two control units in the matched set}, we may have two treatment-minus-control differences, $Y_{11}$ and $Y_{12}$, such that if the first unit is treated, then $Z_{11} = 1$ and treated-minus-control differences are $R^l_{11} - R^l_{12}$ and $R^l_{11} - R^l_{13}$, while if the second unit is treated, then $Z_{12} = 1$ and treated-minus-control differences are $R^l_{12} - R^l_{11}$ and $R^l_{12} - R^l_{13}$, and so on. In general, if $J_i$ is the {unique} unit treated in the $i$th matched set with $J_i = \sum_{j=1}^{n_i} jZ_{ij}$, then the $n_i - 1$ treated-minus-control differences in set $i$ for the $l$th outcome are $Y^l_{ik} =R^l_{i,J_i} - R^l_{ik}$, with ${1\le k\le n_i, k\ne J_i}$. {Analogously, if  $J_i$ is the {unique} control unit in the $i$th matched set, with $J_i = \sum_{j=1}^{n_i} j(1-Z_{ij})$, then the $n_i-1$ treated-minus control differences in set $i$ for the $l$th outcome are $Y^l_{ik} = R^l_{ik} - R^l_{i,J_i}$, with $k\in \{1,\cdots, n_i\}\backslash J_i$. The \texttt{sensitivityfull} package in \texttt{R} implements tests and sensitivity analysis for observational studies that conduct full matching in this manner.}

Now, one can look the treated-minus-control differences and compare them with the control-minus-control {or treatment-minus-treatment} differences in the matched set, and thus assign a significant departure from the null distribution under no-unmeasured confounding to be an evidence in favor of the alternative. Define $N_{i,-J_i} = \{1\le k\le n_i, k\ne J_i\}$ {where $J_i$ is the unique treated or unique control unit in a matched set. Hence $N_{i,-J_i}$ contains all the like elements in a matched set, all treated or all control.} %
One can thus consider the generalized test-statistic
$$T_g^l(\bm Z,\bm R^l) = \dfrac{\sum_{i=1}^I\sum_{k\in N_{i,-J_i}}\sgn(Y_{ik}^l)q^l_{ik}}{\sum_{i=1}^I\sum_{k\in N_{i,-J_i}} q_{ik}^l},$$
where $q_{ik}^l$ are again appropriately chosen functions of the treated-minus-control differences and the statistic has been normalized. One may refer to \cite{rosenbaum2007sensitivity} Section 4.2 for a detailed discussion. Unlike the matched pairs setting, there does not exist a bounding random variable $T_{\bar\Gamma}^l$ in the context of Model \ref{rosen-model} that holds in finite samples. However, one can obtain a random variable $T^l_{\bar\Gamma, g}$ such that, conditional on $\F$ and $\Z$, %
\begin{equation}
    \label{multi-ass}
    \begin{aligned}
    \dfrac{\sqrt I(T^l_{\bar\Gamma, g} - a^l_{\Gamma})}{b^l_{\Gamma}}&\sim \mathcal{N}(0,1) \quad \text{and} \quad
    \P(T_{l,g}\ge t|\F,\Z)\le \P(T^l_{\bar\Gamma, g}\ge t|\F,\Z) \text{ as }I\to\infty,%
    \end{aligned}
\end{equation}
for some appropriate choice of $a^l_{\Gamma}$ and $b^l_{\Gamma}$ (\cite{gastwirth2000asymptotic}, \cite{rosenbaum2007sensitivity}) when $n_i\le p$ for all $i$ and regularity conditions on $q_i$s. Now, suppose one conducts the sensitivity analysis using Equation \ref{multi-ass}, then the sensitivity value would satisfy %
\begin{equation} \label{rep1}
    T^l_g - a^l_{\Gamma^*_{\alpha}(\bm Z,\bm R^l)} = \Phi^{-1}(1-\alpha_l)\dfrac{b_{\Gamma^*_{\alpha}(\bm Z,\bm R^l)}}{\sqrt I} + o_p\left(\dfrac{1}{\sqrt I}\right)%
\end{equation}

Then, suppose the true distribution of $T^l_{g}$, akin to Assumption \ref{alt-clt}, admits an asymptotic distribution of the form %
\begin{equation} \label{alt-clt-multi}
\sqrt{I}(T^l_g - \mu_{Fg}^l)\overset{d}{\to} \mathcal{N}(0,(\sigma_{F,g}^l)^2)%
\end{equation}
Then, we can have a result analogous to \ref{Qing-main}.
\begin{theorem} \label{sens-multi}
    Suppose Equations \ref{multi-ass} and \ref{alt-clt-multi} hold. %
    Also, suppose that the mapping $\Gamma\mapsto a_{\Gamma}^l$ is continuously differentiable at a neighborhood of $\tilde \Gamma^l$ and a has non-zero derivative at $\tilde \Gamma^l$, where $\tilde\Gamma^l$ is the design sensitivity of the $l$th outcome. Then,%
    $$\sqrt I(\Gamma^*_{\alpha}(\bm Z,\bm R^l) - \tilde\Gamma^{l}) \overset{d}{\to}\mathcal{N}(-\Phi^{-1}(1-\alpha_l) \mu^g_{\tilde\Gamma^l}, (\sigma^{g}_{\tilde\Gamma^l})^2)%
    $$
    for some constants $\mu^g_{\tilde\Gamma^{l}}$, $\sigma^g_{\tilde\Gamma^l}$.
\end{theorem}

We can use the same strategy as in Section \ref{sec:ourmethod} using Theorem \ref{sens-multi}, but the expressions for $\mu_{\tilde\Gamma}^{l,g}$ are again fairly complicated. However, this time we can estimate both the mean and variance of the standardized asymptotic distribution using the bootstrap. 

\begin{method}
   \label{bootstrap-multi}
   {{(Sens-Val-Full-Matching)}}. Consider the approach of sample splitting for hypothesis screening. To control for bias at level $\Gamma_{\c}$ in the analysis sample when using multiple controls in matched sets, for each outcome $l$: 
   \begin{itemize}
        \item Compute $\Gamma^*_{\alpha_{\p}}(\bm Z_{\p}, \bm R^l_{\p})$, the sensitivity value for the $l^{\text{th}}$ outcome at level $\alpha_{\p}$.
        \item Take $B$ bootstrap samples of $I_{\p}$ matched sets from $(\bm Z_{\p}, R^l_{\p})$ and calculate sensitivity values $\{\Gamma^*_{\alpha_{\p},b}(\bm Z_{\p,b}, \bm R^l_{\p,b})\}_{b \in \{1, \cdots, B\}}$. Compute %
        $$(\hat\sigma^l_{F,g})^2 = I\widehat\V(\Gamma^*_{\alpha_{\p},b}(\bm Z_{\p,b}, \bm R^l_{\p,b})).%
        $$ Also, compute $\Gamma^*_{\alpha_{\p}, \text{boot}}(\bm Z_{\p}, \bm R^l_{\p})$, the sample mean of the bootstrap samples, and obtain %
        $$\hat\mu^{l,g}_{\tilde\Gamma} = \sqrt{I}(\Gamma^*_{\alpha_{\p}}(\bm Z_{\p}, \bm R^l_{\p}) - \Gamma^*_{\alpha_{\p}, \text{boot}}(\bm Z_{\p}, \bm R^l_{\p}))%
        $$
        \item Test for the outcomes $H_{0l}:l\in S^\text{Sens-Val-Multiple-Controls}_{\p}(\Gamma_{\c}, \alpha_{\p})$, 
        is given by
        \begin{align*} \left\{l\in[L]: \Gamma^{*}_{\alpha_{\p}}(\bm Z,\bm R^l_{\p}) + \dfrac{\widehat\mu_{\tilde\Gamma}^{l,g}}{\sqrt I}\left[\dfrac{\Phi^{-1}(1-\alpha_{\p})}{\sqrt r} - \dfrac{\Phi^{-1}(1-\alpha_l)}{\sqrt {1-r}}\right]>\right.\\ \left. \kappa_{\c} - \dfrac{\hat\sigma^l_{F,g}\Phi^{-1}(1-\alpha_{\cv})}{\sqrt{Ir(1-r)}} \right\}
        \end{align*}
        where $[L] = \{1,\cdots,L\}$, and $\{\alpha_l\}_1^L$ and $\alpha_{\cv}$ are hyperparameters to be determined by the researcher.
    \end{itemize}
\end{method}

\begin{theorem} \label{ourmain-multi}
     If $\hat\mu_F^{l,g}$ and $\hat\sigma_F^{l,g}$ are consistent for their corresponding parameters and the conditions for Theorem \ref{sens-multi} hold, then
    \begin{align*}
    \P\left(\Gamma^*_{\alpha_l}(\bm Z_{\a}, \bm R^l_{\a}) \ge \Gamma^{*}_{\alpha_{\p}}(\bm Z_{\p},\bm R^l_{\p}) + \dfrac{\widehat\mu_{\tilde\Gamma}^{l,g}}{\sqrt I}\left[\dfrac{\Phi^{-1}(1-\alpha_{\p})}{\sqrt r} - \dfrac{\Phi^{-1}(1-\alpha_l)}{\sqrt {1-r}}\right] \right. &\\ \left.+ \dfrac{\hat\sigma^l_{F,g}\Phi^{-1}(1-\alpha_{\cv})}{\sqrt{Ir(1-r)}} \right) = (1 - \alpha_{\cv}) + o(1)
    \end{align*}
\end{theorem}
The proof of Theorem \ref{ourmain-multi} is \textit{mutatis mutandis} identical to the proofs of Theorem \ref{ourmain1} and Corollary \ref{ourcor1}, and is thus omitted. Theorem \ref{ourmain-multi} provides a way to select outcomes from the planning sample, even under choice of multiple controls.

\section{Choice of Hyperparameters}

\subsection{Hyperparameter search}\label{hypersearch}

During the experiment run in Section \ref{sec:ourmethod}, we set $\{\alpha_l\}_1^L = \alpha/L$ and $\alpha_{\cv}=\alpha$ for the Sens-Val procedure. Our theoretical results presume that $\alpha_l$ is known \textit{a priori} and the sensitivity values for each $l$ are calculated under this belief. However, if the researcher uses a Bonferroni correction, i.e., $\alpha_l = \alpha\mathbbm{1}(l\in S^\text{Sens-Val}_{\p})/|S^\text{Sens-Val}_{\p}|$, then $\alpha_l$s are implicit and the outcomes are generally not  selected optimally for the criteria they would have to face in the analysis sample. In this regard, we can implement a dynamic $\alpha_l$ framework as follows:
    \begin{itemize}%
        \item Set $\alpha_l^0 = \alpha/L$, for each $l$.
        \item For $n\ge 1$, compute $S^\text{Sens-Val}_{\p}(\Gamma_{\c}, \alpha_{\p})^{n-1}$ at $\alpha_l = \alpha_l^{n-1}$.
        \item Update $\alpha_l^n = \alpha/|S^\text{Sens-Val}_{\p}(\Gamma_{\c}, \alpha_{\p})^{n-1}|$.
        \item Continue till convergence of $\{\alpha_l^n\}_{n\ge 1}$ for all $1\le l\le L$, and report $S^\text{Sens-Val}_{\p}(\Gamma_{\c},\alpha_{\p})$ as $S^\text{Sens-Val}_{\p}(\Gamma_{\c},\alpha_{\p})$ at the converged values of $\{\alpha^n_l\}_{n\ge 1}$ for each $l$.
    \end{itemize}

Although this procedure does not necessarily have the same statistical guarantees with respect to the coverage of the sensitivity values in the analysis sample, we find the dynamic $\alpha_l$ framework tends to demonstrate slightly enhanced performance in simulations, particularly in settings with a sparse number of true signals. Intuitively, we are selecting outcomes based on criteria as stringent as they would actually face in the analysis sample, and the selection bias resulting from the dynamic choice of $\alpha_l$s turns out to be insignificant. We can safely do this since Proposition \ref{validity} ensures FWER control in the analysis sample.

The Sens-Val method is relatively stable with the choice of $\alpha_{\p}$, while the researcher can tune $\alpha_{\cv}$ based on how liberal or stringent she wishes to be. We set $\alpha_{\p} = \alpha_{\cv} = \alpha$ throughout the main text and conduct simulations and data analysis with dynamic $\alpha_l$s.

{
\subsection{Performance for different values of $\alpha_{\cv}$}
The choice of $\alpha_{\cv}$ is an interesting, if perhaps surprisingly, not the most important consideration to the performance of our Sens-Val procedure. To illustrate, we conduct simulations that emulate the data-inspired setting described in Section \ref{results} of our manuscript and compare the performance of our approach for $r \in \{1/10, 1/5, 1/3\}$ and $\Gamma_{\text{data}} = \Gamma_{\text{con}} \in \{1.25, 2.50\}$. 

\begin{figure}[!h]
\centering
\includegraphics[width = \textwidth-4em]{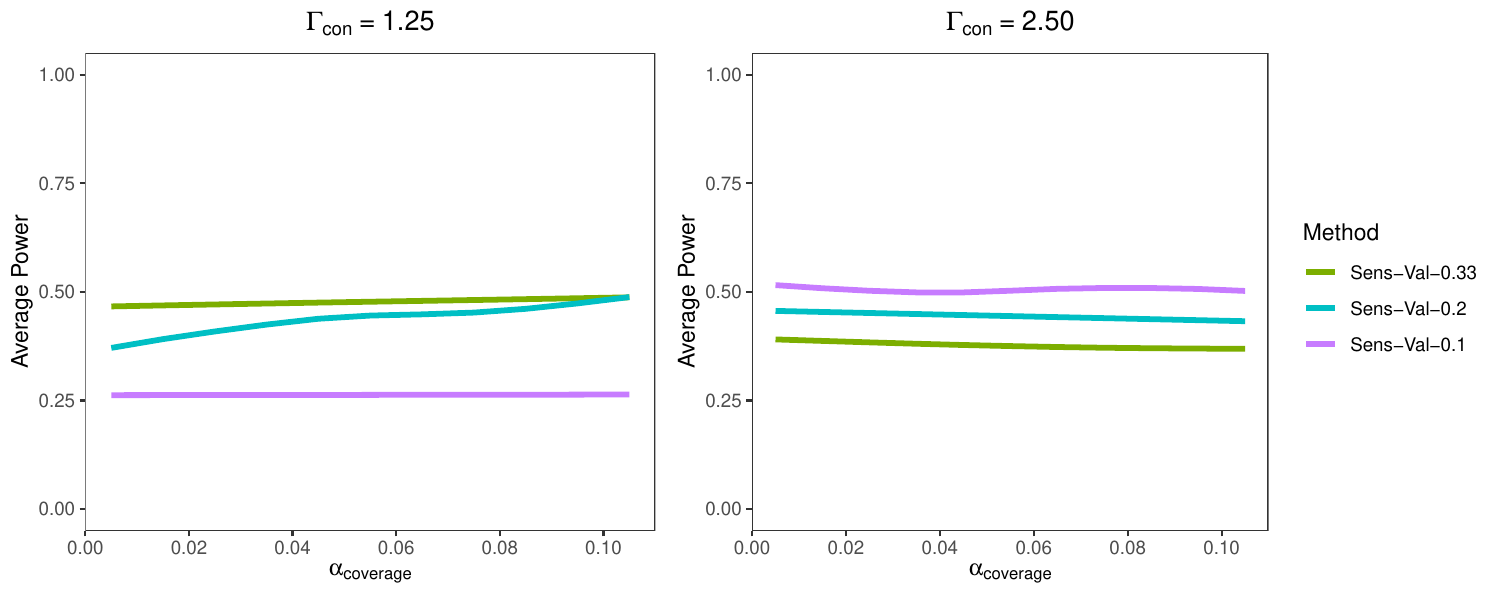}
\caption{Simulation results for the data-inspired UC setting varying $\alpha_{\cv}$, with multiple values of $r$ and $\Gamma$.}
\label{fig:alphacoverage-plots}
\end{figure}

Recall that in our procedure, the parameter $\alpha_{\cv}$ relates to the length of predictive intervals constructed on the analysis sample, while the bias correction term (involving hyperparameters $\alpha_{\p}$ and $\alpha_{l}$) determines where to center these intervals (see Theorem \ref{ourmain1}). By Corollary \ref{ourcor1}, we obtain $(1-\alpha_{\cv})$ predictive interval for the sensitivity value of each outcome when tested on the analysis sample at level $\alpha_l$. 
However, we observe empirically that the performance of our method only fluctuates mildly with respect to changes in the value of $\alpha_{\cv}$, across various choices of split-sampling proportions $r$ and levels of unmeasured confounding $\Gamma_{\text{con}}$. Like many settings that seek to infer causal effects, our intuition is that properly accounting for bias (selection of $\alpha_{\p}$ and $\alpha_{l}$) is indeed the dominant consideration (more so than variance and choice of $\alpha_{\cv}$, in contrast with many usual statistical inference procedures) in the performance of our method.}

{\section{Additional approximations for moderate number of matched pairs}}
{\cite{zhao2018sensitivity} highlights that while Theorem \ref{ourmain1} based on Equation \ref{Qing-main} assumes $\Phi^{-1}(1-\alpha)$ to be fixed, which allows us to asymptotically eliminate a lot of the terms, ignoring such terms may be sub-optimal in finite samples. He points out that at $I=50$, $\alpha = 0.05$, $\sigma_q^2 \Phi^{-1}(1-\alpha)$ (suppressing the superscript $l$) for the Wilcoxon signed rank statistic is non-negligible, and hence one can hope to improve power if such terms are considered. Thus one may be interested in selecting outcomes when $I$ is relatively small-enough, in comparison to $\Phi^{-1}(1-\alpha)$. In this section we shall discuss how such a method may be obtained.}

{We shall begin by the following approximation equation, assuming Equation \ref{alt-clt}, where we let $\sigma_q\Phi^{-1}(1-\alpha_{\p})/\sqrt{I_{\p}} =:\sqrt{\eta_{\p}}$, and treat $\eta_{\p}$ as fixed. $\eta_l$ is defined and treated analogously, replacing $\alpha_{\p}$ by $\alpha_l$.
\begin{equation}
    \label{approx}
    \sqrt{I_{\p}}\left(\kappa_{\alpha_{\p}} +\sqrt{\eta_{\p}\kappa_{\alpha_{\p}}(1-\kappa_{\alpha_{\p}})} -\mu_F\right) \approx \mathcal{N}(0,\sigma_F^2).
\end{equation}
Thus, in this regime, although $\kappa_{\alpha_{\p}}$ is not consistent for $\mu_F$, $\hat\mu_F := \kappa_{\alpha_{\p}} + \sqrt{\eta_{\p}\kappa_{\alpha_{\p}}(1-\kappa_{\alpha_{\p}})}$ is so, at a rate of $O_p(1/\sqrt{I_{\p}})$. Now we use the result for $\kappa_{\alpha_l}$ in this regime from \cite{zhao2018sensitivity}, which states that 
\begin{align*}
        &\sqrt{I_{\a}}\left(\kappa_{\alpha_{l}} - \mu_F + \dfrac{(2\mu_F-1)\eta_l + \sqrt{4\eta_l\mu_F(1-\mu_F)+\eta_l^2}}{2(1+\eta_l)}\right)\\ &\approx \mathcal{N}\left(0,\dfrac{\sigma_F^2}{(1+\eta_l)^2}\left(1+\dfrac{\eta(2\mu_F-1)}{\sqrt{4\eta_l\mu_F(1-\mu_F)+\eta_l^2}}\right)^2\right).
\end{align*}
or for ease of writing, $\sqrt{I}(\kappa_{\alpha_l} - f(\mu_F)) \approx \mathcal{N}(0,\sigma_F^2 g(\mu_F)^2)$, where 

$f(\mu_F) =  \mu_F -\dfrac{(2\mu_F-1)\eta_l + \sqrt{4\eta_l\mu_F(1-\mu_F)+\eta_l^2}}{2(1+\eta_l)}$ and

$g(\mu_F) = \dfrac{1}{(1+\eta_l)}\left(1+\dfrac{\eta(2\mu_F-1)}{\sqrt{4\eta_l\mu_F(1-\mu_F)+\eta_l^2}}\right)$. 
Thus, one may obtain 
\begin{align*}
    \sqrt{I}(\kappa_{\alpha_l} - f(\hat\mu_F)) &= \sqrt{I}(\kappa_{\alpha_l} - f(\mu_F)) - \sqrt{I}(f(\hat\mu_F) - f(\mu_F)) \\
     &= \sqrt{1-r}\sqrt{I_{\a}}(\kappa_{\alpha_l} - f(\mu_F)) - \sqrt{r}\sqrt{I_{\p}}(\hat\mu_F - \mu_F)f'(\mu_F) + o_p(1)\\
     &\approx \mathcal{N}\left(0,\sigma_F^2((1-r)g(\mu_F)^2 + rf'(\mu_F)^2)\right).
\end{align*}
One can thus use the above equation to obtain a selection method when $\sqrt{I}$ is moderate-sized in comparison to $\alpha_{\p}$ and $\alpha_l.$}

\vspace{3em}
{\begin{method}
    \label{appmethod}
    {(Approximate-Sens-Val)} Consider the approach of sample splitting for hypothesis screening and suppose we want to control for bias at level $\Gamma_{\c}$ in the analysis sample. Then for each outcome $l$
    \begin{itemize}
        \item Repeat the first two steps in Method \ref{bootstrap}. Compute $\hat\mu_F =  \kappa_{\alpha_{\p}} + \sqrt{\eta_{\p}\kappa_{\alpha_{\p}}(1-\kappa_{\alpha_{\p}})}$
        \item Test for the outcomes $H_{0l}:l\in S_{\p}^{\text{arcsin-SensVal}}(\Gamma_{\c},\alpha_{\p})$, where 
        \begin{align*} S_{\p}^{\text{approx-SensVal}}(&\Gamma_{\c},\alpha_{\p}) = \{l\in [L]: 
        \kappa_{\c} < \\&f(\hat\mu_F) + \dfrac{\hat\sigma_F^2((1-r)g(\hat\mu_F)^2 + rf'(\hat\mu_F)^2)}{\sqrt{I}}\Phi^{-1}(1-\alpha_{\cv})
        \},
        \end{align*}
        where $f$ and $g$ are as defined above, and $\{\alpha_l\}_{l=1}^L$ and $\alpha_{\cv}$ are hyperparameters to be determined by the researcher.
    \end{itemize}
\end{method}}

{{Proof of Equation \eqref{approx}}: We begin with the large sample approximation $\sqrt{I}(T-\kappa_\alpha) = \sqrt{\kappa(1-\kappa)}\sigma_q(z_\alpha + o_p(1)) = \sqrt{I}\eta \sqrt{\kappa_\alpha(1-\kappa_\alpha)} + o(1)$. 
Thus, by Equation \ref{alt-clt},
\begin{align*}
    \sqrt{I}\left(\kappa_\alpha +\eta\sqrt{\kappa_\alpha(1-\kappa_\alpha)} -\mu_F\right) = \sqrt{I}(T-\mu_F) + o_p(1) \approx \mathcal{N}(0,\sigma_F^2).
\end{align*}}

\section{Additional Design Considerations}

{
\subsection{Choice of split-sampling ratio}
The choice of split-sampling ratio is an important parameter, yet there is currently no agreed upon standard for doing so. \cite{heller2009split} conducted simulations assessing the power of splitting for $r \in \{1/10, 1/6, 1/3\}$ and found no ratio to be uniformly most powerful across data generating processes. \cite{zhang2011using} studied the effect of hospital closures on mothers and their newborns using roughly $266,000$ mothers in roughly $133,000$ matched pairs, and set $r=1/10$. A later study by \cite{lee2021discovering} examined the impact of long-term exposure to air pollution on mortality with more than $1.61$ million individuals in just over $110,000$ matched pairs, taking $r = 1/4$. They also conducted a simulation to assess the effects of varying the splitting ratio on the power of their analysis, highlighting the trade-off between exploration and confirmation. }

{To get a better sense of how our method performs based on the choice of split-sampling ratio $r$, we conduct simulations which emulate the data-inspired setting described in Section \ref{results} of the main text and compare the performance of our approach for $r \in \{1/10, 1/5, 1/3\}$ and $\Gamma_{\text{data}} = \Gamma_{\text{con}} \in \{1.25, 2.50\}$. %

\begin{figure}[h]
\centering
\includegraphics[width = \textwidth-4em]{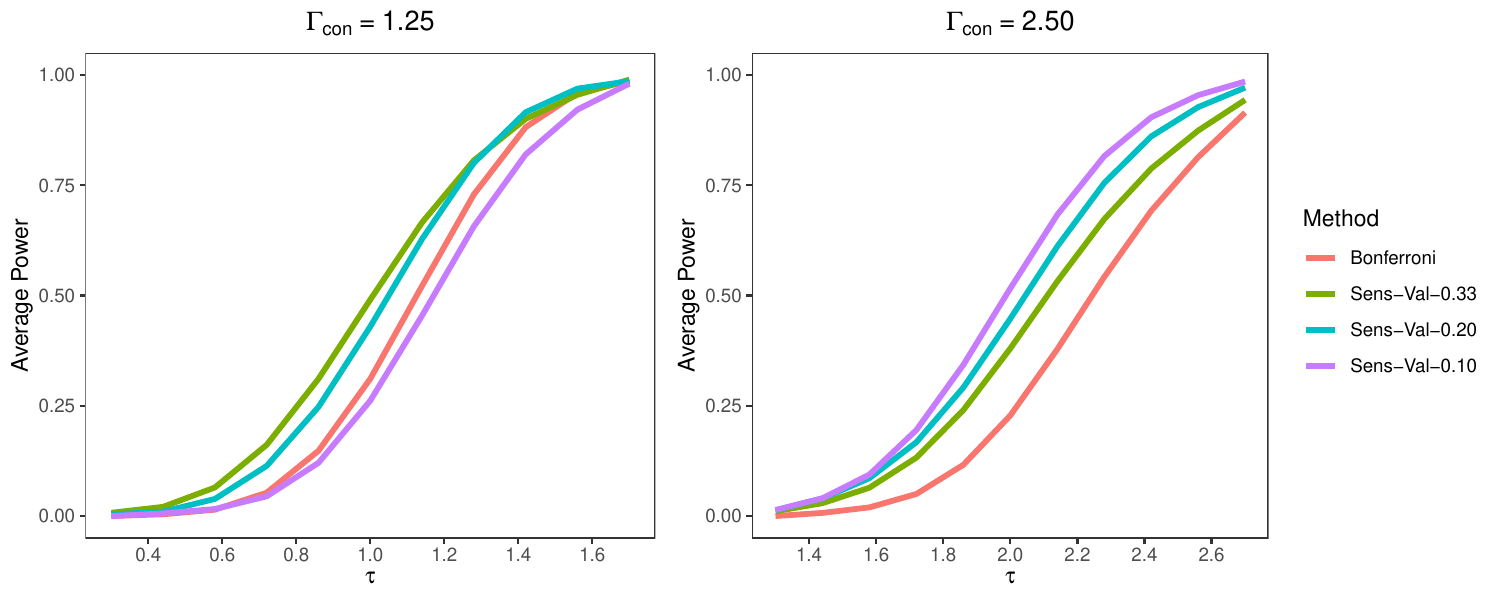}
\caption{Simulation results for the data-inspired UC setting with multiple values of $r$ and $\Gamma$.}
\label{fig:r-plots}
\end{figure} 

We observe that our Sens-Val method performs well with larger values of $r$ when controlling for mild levels of potential unmeasured confounding (i.e., $\Gamma_{\c}=1.25$) and smaller values of $r$ when controlling for more substantial levels of potential unmeasured confounding (i.e., $\Gamma_{\c}=2.50$). Our intuition is that when $\Gamma_{\c}$ is small, it is relatively challenging for a filter to distinguish true signals (that is, those with signals above the design sensitivity threshold) from noise, and thus a larger sample size is required to establish a successful filtering. On the other hand, when $\Gamma_{\c}$ is large, the non-null outcomes have a stronger signal, so the separation between nulls and non-nulls is larger and our method requires less sample size to detect null outcomes and filter them out. 

Our pick of $r=0.20$ in the manuscript seems to be a good choice across the simulation settings we considered, which were in part inspired by the setting for the Bangladesh Flood study.
}

\subsection{Choice of statistics}

Rosenbaum advocates for catering design with the intent of maximizing design sensitivity. \cite{zhao2018sensitivity} applies this idea to the selection of test statistics with the sensitivity value. He considers the generalized score statistic where $q_i^l = \psi^l(\text{rank}(Y_i^l))/(I+1)$ and obtains a formula for the asymptotic mean and variance in Equation \ref{Qing-main} based on the outcome density and score function $\psi^l$ which can vary by outcome. However, his analysis relies only on the mean, as the variance parameter is quite involved to calculate. It also requires prior knowledge of the outcome distributions.

Due to our use of sample splitting, we can bypass excess considerations and directly compare test statistics on the planning sample. For instance, suppose the $l^{\text{th}}$ outcome has $\{\psi^l_k\}_{1\le k\le p_l}$ candidate choices to be used. Although we have framed Sens-Val as a procedure for outcome selection, one can redefine an outcome as its $(l,k)$ pair, $1\le k\le p_l$, $1\le l\le L$, and choose the (outcome, statistic) pair suggested by the method. In case of a tie among the statistics, one can simply choose the statistic with the higher sensitivity value. Such a choice bypasses the theoretical considerations of \cite{zhao2018sensitivity}, circumvents dealing with variances $\sigma_F^l$, and provides a data-driven means for selection of outcomes and test statistics. Moreover, it is driven by the quantiles of the asymptotic distribution of the sensitivity value, adhering to the suggestions by \cite{zhao2018sensitivity} and augmenting it with the possibility that quantiles used in planning and analysis samples may differ.

We illustrate the potential benefits of this idea as follows. Consider a setting with $N=200$ subjects randomly assigned to treatment or control with probability $1/2$ and $L=20$ possible outcomes, among which there exist four true signals. Suppose the control potential outcomes $r^l_{C_{ij}}$ are all generated independently from either the standard Normal distribution, standard logistic distribution, or a $t$-distribution with four degrees of freedom. Treated units amongst the true signals equal their control potential outcome plus a constant treatment effect of $\tau$, which is specified for each distribution. This setup allows us to assess the performance of our method for various types of data. Using a planning sample proportion of $0.20$, we evaluate the average power of our Sens-Val method across 100 simulations for various elements in the class of $U$-statistics proposed by \cite{rosenbaum2011new, rosenbaum2014weighted}, which are indexed by parameters $(m, \underline{m}, \overline{m})$. 

For a fixed outcome $l$, we let \( h({y}) \) denote a function of \( m \) variables that quantifies the positive differences among the order statistics between \( |y|_{(\underline{m})} \) and \(|y|_{(\overline{m})} \). The corresponding \( U \)-statistic can then be taken as \( T = \binom{I}{m}^{-1} \sum_{|\mathcal{I}| = m} h(Y_{\mathcal{I}}) \) and written as signed score form \ref{pairstats}. In the absence of ties:
\begin{align*}
    q_i^l &= \binom{I}{m}^{-1} \sum_{k=\underline{m}}^{\overline{m}} \binom{\text{rank}(|Y_i^l|) - 1}{k - 1} \binom{I - \text{rank}(|Y_i^l|)}{m - k}
    \approx I^{-1} \sum_{k=\underline{m}}^{\overline{m}} k \binom{m}{k} p^{k-1} (1 - p)^{m-k},
\end{align*}
where \( p = \text{rank}(|Y_i^l|)/I \) (\cite{rosenbaum2011new}, Section 3.1). Note that the choice \( (m, \underline{m}, \overline{m})= (2, 2, 2) \) closely approximates Wilcoxon’s signed rank statistic (\cite{pratt1981concepts}).

We observe the simulated average power among these statistics in Table \ref{tab:many-stats} below. In terms of performance, the best statistic to use often changes based on the simulation setting and data generating process. This shows that it can often be beneficial in terms of detection power to compare statistics to use during the planning stage.

\begin{table}[!ht]
\centering
\caption{Simulated average power with $N=200$ subjects of a $0.05$ level analysis for several values of $\Gamma$. The power is the proportion of simulations where the test correctly rejects the null hypothesis when it is false, performed at the given value of $\Gamma$. Each situation is replicated 100 times. Scores above the horizontal line are monotone increasing scores, while those below are descending scores.}
\label{tab:many-stats}
\scriptsize
\resizebox{\textwidth}{!}{\begin{tabular}{lccc|ccc|ccc}
\toprule
& \multicolumn{3}{c}{Normal(0,1)} & \multicolumn{3}{c}{Logistic(0,1)} & \multicolumn{3}{c}{$t$ with 4 d.f.} \\
& \multicolumn{3}{c}{$\tau = 1$} & \multicolumn{3}{c}{$\tau = 1.5$} & \multicolumn{3}{c}{$\tau = 1.5$} \\
\cmidrule(lr){2-4} \cmidrule(lr){5-7} \cmidrule(lr){8-10}
$(m, \underline{m}, \overline{m})$ & $\Gamma=1.5$ & $\Gamma=2.5$ & $\Gamma=3.5$ & $\Gamma=1.5$ & $\Gamma=2.5$ & $\Gamma=3.5$ & $\Gamma=1.5$ & $\Gamma=2.5$ & $\Gamma=3.5$ \\
\midrule
(2,2,2)    & 0.855 & 0.565 & 0.150 & \textbf{0.648} & 0.200 & 0.048 & 0.390 & 0.113 & 0.013 \\
(5,4,5)    & 0.843 & 0.555 & 0.183 & 0.620 & 0.215 & 0.055 & 0.368 & 0.110 & 0.015 \\
(8,7,8)    & 0.778 & 0.463 & 0.165 & 0.495 & 0.160 & 0.043 & 0.243 & 0.068 & 0.015 \\
(8,5,8)    & \textbf{0.860} & 0.573 & 0.193 & 0.643 & \textbf{0.225} & \textbf{0.058} & 0.413 & 0.133 & 0.018 \\
(20,16,20) & 0.753 & 0.445 & 0.180 & 0.463 & 0.170 & 0.033 & 0.225 & 0.058 & 0.013 \\
(20,12,20) & 0.848 & \textbf{0.575} & \textbf{0.223} & 0.618 & 0.223 & \textbf{0.058} & 0.398 & 0.130 & 0.020 \\
\midrule
(8,7,7)    & 0.723 & 0.438 & 0.178 & 0.500 & 0.173 & 0.045 & 0.380 & 0.108 & \textbf{0.025} \\
(8,5,7)    & 0.753 & 0.423 & 0.115 & 0.563 & 0.168 & 0.040 & \textbf{0.515} & 0.128 & 0.023 \\
(20,16,19) & 0.653 & 0.368 & 0.170 & 0.425 & 0.153 & 0.053 & 0.260 & 0.078 & \textbf{0.025} \\
(20,12,19) & 0.760 & 0.488 & 0.173 & 0.548 & 0.185 & 0.050 & 0.483 & \textbf{0.143} & 0.023 \\
\bottomrule
\end{tabular}}
\end{table}

\subsection{Choice of sub-populations} %
\cite{zhao2018sensitivity} considers the choice of sub-populations to be an important aspect of the study design and explains the trade-off of using sub-populations instead of all subgroups observed by \cite{hsu2013effect}. While this explains the phenomenon based on the difference of design sensitivities (and sub-group proportions), the ignorance of design-sensitivity may leave a researcher no better off, unless she employs a pilot sample to estimate the design sensitivities. As noted before, estimation of design sensitivity is not only challenging, but also not quite the goal of the researcher who has to run the tests on the analysis sample. Sens-Val solves the problem more pragmatically -- one may consider the effect on all possible sub-populations as different outcomes, and use our method to select the sub-populations which have a reasonable chance to be significant in the analysis sample.

We demonstrate this advantage with a simulation. Consider a setting with $N=2000$ subjects randomly assigned to treatment or control with probability $1/2$ and $L=10$ possible outcomes, among which there exist three true signals. We use a planning sample proportion of $0.20$ to divide control/treated pairs into discrete planning and analysis samples. We then take $K=5$ discrete sub-populations, where control/treated pairs within the planning and analysis samples are assigned to sub-population $k$ with probability $\pi_k = (0.10,0.15,0.20,0.20,0.35).$ Suppose the control potential outcomes $r^l_{C_{ij}}$ are all generated independently from the standard Normal distribution and treated potential outcomes are taken as:
\begin{equation*}
r^l_{T_{ij}} =
\begin{cases}
r^l_{C_{ij}} + \frac{2}{3}, & \text{if } l = 1 \text{ or } (l = 2 \text{ and } k \in {1,2,3}) \text{ or } (l = 3 \text{ and } k \in {1,2});\\
r^l_{C_{ij}} + \frac{1}{3}, & \text{if } l = 3 \text{ and } k \in {3,4};\\
r^l_{C_{ij}}, & \text{otherwise}.
\end{cases}
\end{equation*}
The Sens-Val method selects any promising sub-populations in the planning sample to test in the latter stage of the pipeline. We conduct our Sens-Val method and Bonferroni correction across 1,000 simulations at $\alpha=0.05$ and $\Gamma_{\c} = 2$, and display our simulated performance in Table \ref{tab:subpops} below. We see that Sens-Val does a very good job choosing sub-populations during the planning stage, which yields notable benefits in terms of detection power.

\begin{table}[!ht]
\centering
\caption{Simulation results whereas each situation is replicated 1,000 times. The left-hand-side shows the proportion of simulations in which an outcome/sub-population pair is selected by Sens-Val in the planning sample. The right-hand-side gives the frequency for which each method rejects a certain outcome, along with an average true positive rate for each method. Bolded numbers depict scenarios with non-zero treatment effects.}
\label{tab:subpops}
\small
\begin{tabular}{cccccc|cc}
\toprule
\multicolumn{6}{c}{{Proportion Tested}} & \multicolumn{2}{c}{{Proportion Rejected}} \\
\cmidrule(r){1-6} \cmidrule(l){7-8}
& \multicolumn{5}{c}{{Sub-Population}} & \multicolumn{2}{|c}{{Method}} \\
\cmidrule(r){2-6} \cmidrule(l){7-8}
{Outcome} & {1} & {2} & {3} & {4} & {5} & {Sens-Val} & {Bonferroni} \\
\midrule
1 & \cellcolor{gray!50}\bf 0.999 & \cellcolor{gray!50}\bf 1.000 & \cellcolor{gray!50}\bf 1.000 & \cellcolor{gray!50}\bf 1.000 & \cellcolor{gray!50}\bf 1.000 & 0.923 & 0.917 \\
2 & \cellcolor{gray!50}\bf 1.000 & \cellcolor{gray!50}\bf 1.000 & \cellcolor{gray!50}\bf 1.000 & 0.022 & 0.001 & 0.639 & 0.001 \\
3 & \cellcolor{gray!50}\bf 0.999 & \cellcolor{gray!50}\bf 0.999 & \cellcolor{gray!25}$\bf 0.297$ & \cellcolor{gray!25}$\bf 0.470$ & 0.007 & 0.161 & 0.000 \\
4 & 0.009 & 0.006 & 0.004 & 0.010 & 0.005 & 0.000 & 0.000 \\
5 & 0.013 & 0.008 & 0.002 & 0.014 & 0.000 & 0.000 & 0.000 \\
6 & 0.015 & 0.008 & 0.004 & 0.013 & 0.007 & 0.000 & 0.000 \\
7 & 0.010 & 0.007 & 0.009 & 0.015 & 0.007 & 0.000 & 0.000 \\
8 & 0.016 & 0.009 & 0.004 & 0.013 & 0.007 & 0.000 & 0.000 \\
9 & 0.014 & 0.004 & 0.005 & 0.012 & 0.005 & 0.000 & 0.000 \\
10 & 0.007 & 0.009 & 0.007 & 0.017 & 0.005 & 0.000 & 0.000 \\
\cmidrule(l){7-8}
\multicolumn{6}{c}{See that $\textcolor{gray!50}{\qed}$ denotes $\tau =\frac 23$ while $\textcolor{gray!25}{\qed}$ denotes $\tau = \frac{1}{3}$.} & \multicolumn{2}{c}{{True Positive Rate}} \\
\cmidrule(l){7-8}
\multicolumn{6}{c}{} & 0.574 & 0.306 \\
\bottomrule
\end{tabular}
\end{table}

\subsection{Coherence}
An effect is said to be coherent if the observed association between the exposure and outcomes are compatible with the mechanism thought to produce the effects. \cite{rosenbaum1997signed} considers a linear combination of the signed-rank statistics to create a coherent statistic. \cite{ye2022dimensions} extends this idea to all linear combinations using the ideas of \cite{scheffe1953method} and examines the impacts of such choices on the power of the study. \cite{rosenbaum2020combining}, meanwhile, uses an \textit{a priori} guess for the choice of linear combinations obtained from previous data, thereby circumventing the need for multiplicity corrections. We concur that coherence is an important consideration in the design of a study, as a coherent mechanism is less likely to suffer from unmeasured biases. Method \ref{bootstrap} can easily be morphed to impose conditions on the outcomes to test. For example, if an exposure affects an outcome solely through a mediator, the criteria in Method \ref{bootstrap} can be adjusted to include the outcome only if both the impact of exposure on the mediator and the mediator's effect on the outcome meet the specified thresholds. Such restrictions, if known, would be helpful to increase the sensitivity value of the true outcomes while filtering out null or weak signals to enhance discovery.

\subsection{Weighting of outcomes}

Since smaller effects can often arise due to small unattenuated biases, while larger effects can be rendered non-causal only by biases of large magnitudes, one might be encouraged to use weighted statistics. \cite{rosenbaum2014weighted} considers such an approach using weighted M-statistics for superior design sensitivity, but are likely to be driven by the data. In the classical inference literature, \cite{wasserman2006weighted} consider choosing $\alpha_l$s in the approach of sample splitting for hypothesis screening using estimated signal strengths from the planning sample. While the primary concern there is bias rather than efficiency, it may still be useful in a finite sample study tasked with managing both efficiency and bias. Method \ref{bootstrap} used in conjunction with the dynamic $\alpha_l$ framework might lend itself to a middle-ground approach, where the statistics and $\alpha_l$s are suitably driven by the data to obtain enhanced power under bias and moderate sample size. As long as $\alpha_l$s are suitably normalized to control the FWER in the analysis sample, one can experiment via Method \ref{bootstrap} with different choices of the superior statistics based on \cite{rosenbaum2014weighted} and the level thresholds based on signal strengths \cite{wasserman2006weighted} to optimize their power.

{
\section{Additional Simulations}
\subsection{{Changing number of hypotheses}}
To get a better sense of how the Naive method and Bonferroni correction compare as the total number of hypotheses and corresponding proportion of non-nulls change, we conduct additional simulations. We take a total number of subjects $N = 1000$ randomly assigned to treatment or control with probability $1/2$. Each unit records $L$ outcomes, from which there exist $\floor*{L\cdot\text{non-null proportion}}$ outcomes that constitute ``true signals'', comprising the set ${\Upsilon}$. All control potential outcomes $r^l_{C_{ij}}$ are generated independently from the standard Normal distribution and treated potential outcomes are taken as: $r_{T_{ij}}^l = r_{C_{ij}}^l + \mathbbm{1}(l \in {\Upsilon})$. Suppose that we wish to maintain considerable control for possible confounding biases, leading us to set $\Gamma_{\c} = 3.5$. 

We compare the performance of the Naive method and Bonferroni correction for varying number of hypotheses $L$ in both a sparse regime ($5\%$ non-nulls) and a denser regime ($25\%$ non-nulls). We observe in Figure \ref{fig:L-plots} that, for a fixed proportion of non-nulls, as the total number of hypotheses increases, the relative performance of the Naive approach improves compared to Bonferroni correction.

\begin{figure}[!ht]
\centering
\includegraphics[width = \textwidth-4em]{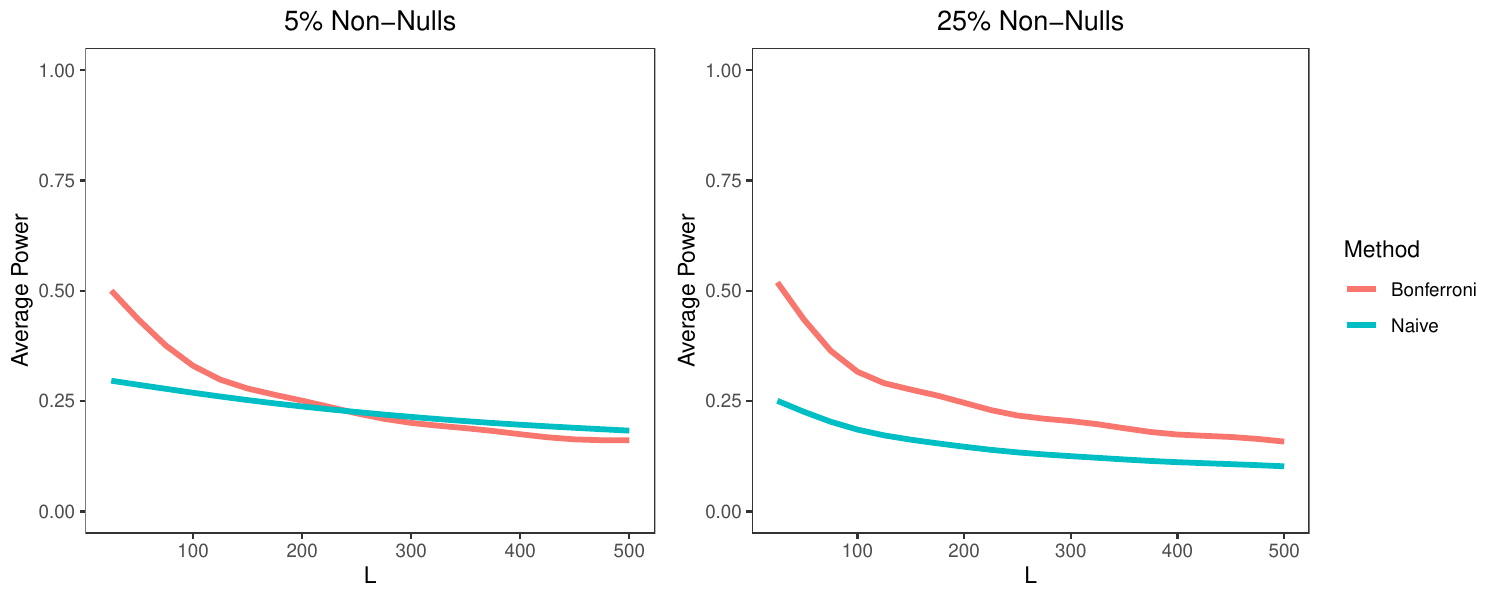}
\caption{Simulation results comparing Naive method and Bonferroni correction with multiple non-null proportions and varying number of hypotheses $L$.}
\label{fig:L-plots}
\end{figure} 

Our intuition is that the average power of both methods degrades as the total number of hypotheses grows and the total number of null outcomes increases, yet the performance of Bonferroni correction declines more sharply since it does not filter out any of the growing number of null outcomes.

As the total number of outcomes increases, the importance of filtering becomes more pronounced since there exist a larger number of null hypotheses, which full-sample Bonferroni must still correct for to maintain valid inference. We also see that the rate of decline of the power curve for the Naive method is smaller in the sparser regime than in the denser regime, since the latter has a greater number of non-null outcomes and so the value of filtering is less pronounced.}

{\subsection{Dense versus sparse signals}}

\begin{figure}[!ht]
        \centering
        \includegraphics[scale=.35]{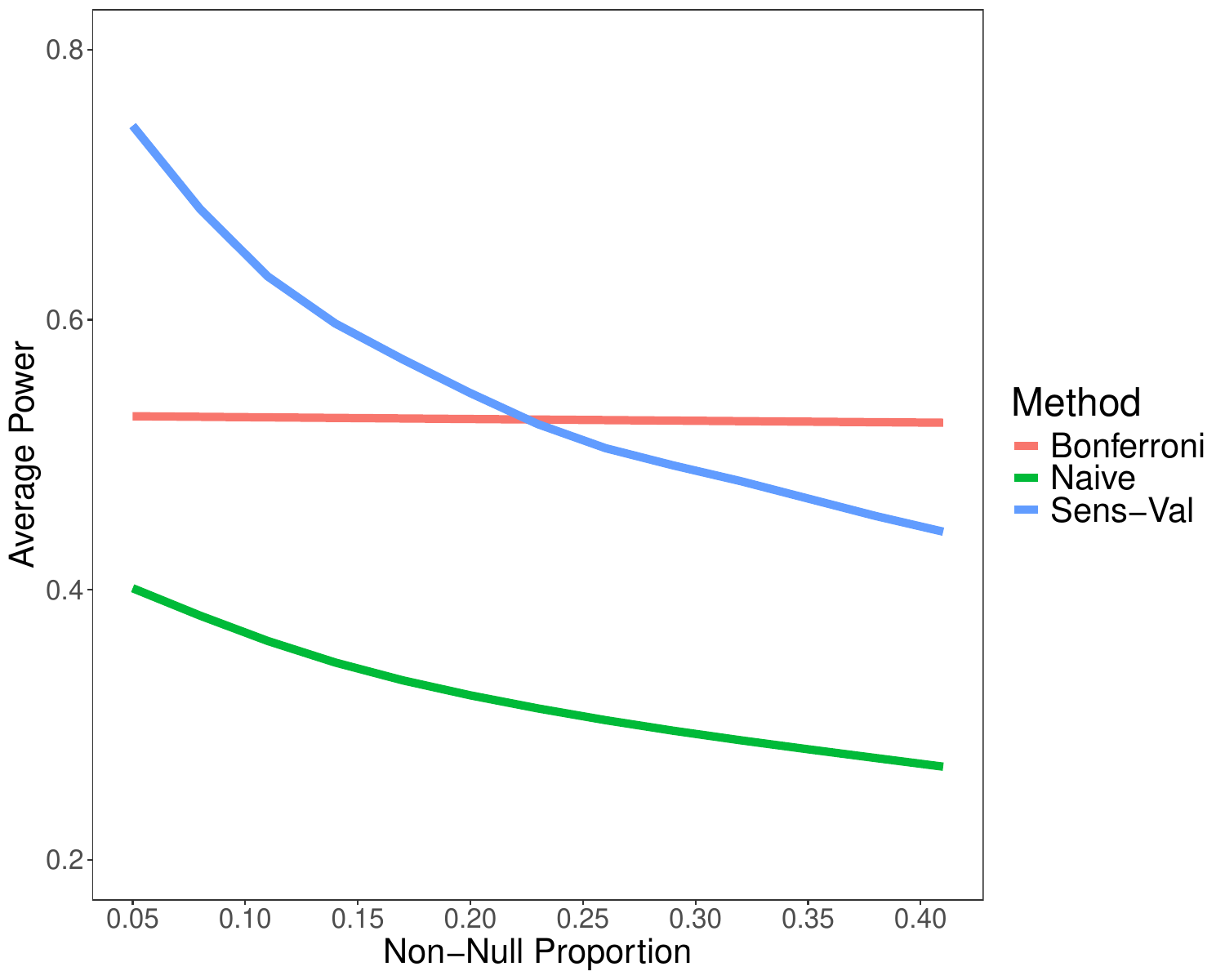}
        \caption{Simulation results comparing full-sample Bonferroni correction, Naive method, and Sens-Val using split samples with varying non-null proportions $|S|/L$.}
        \label{sim-sparse}
    \end{figure}

{
    In this subsection, we compare how Sens-Val performs compared to the Naive approach and full-sample Bonferroni correction in sparse signal regimes. We conduct simulations with a sample size of $N=1500$ subjects randomly assigned treatment with probability $1/2$, and $L=250$ outcomes generated as $R_{T_{ij}} = R_{C_{ij}} + \mathbbm{1}(l\in S)$, $j\in\{1,2\}, 1\le i\le I$, with $R_{C_{ij}}$ distributed as standard normals and $S$ denoting the set of non-null hypotheses. We split the data into a $20\%$ planning sample and $80\%$ analysis sample ($r=0.20$), fix $\Gamma_{\c} = 3.5$, and vary the proportion of non-nulls $|S|/L$ to assess the average power of each procedure which we illustrate in Figure \ref{sim-sparse}.}
    
{We observe that Sens-Val improves uniformly upon the Naive method (as is consistent with Proposition \ref{power} of the manuscript) and outperforms full-sample Bonferroni correction for sparser regimes. When there are many nulls under consideration to be carried on to the analysis sample for formal inference, as is the case in the sparse regime, splitting the data and employing our Sens-Val method is particularly worthwhile since our filter is effective in identifying and screening out null hypotheses. Accordingly, in a sparse regime, it is thus worth the price of reducing our sample size by splitting the data in order to screen out some of the null outcomes via Sens-Val. On the other hand, as the alternative becomes denser, this advantage diminishes and eventually reverses: the utility of an outcome filter is reduced when there are not many null outcomes to be filtered out. 
    Thus in denser regimes, the price of splitting the data and losing sample size begins to outweigh its potential benefits to screen null hypotheses, and the power of the full-sample Bonferroni method starts to exceed that of our Sens-Val approach. However, even in denser regimes, our method maintains the nice feature of enabling the use of exploratory data analysis in designing the study.

    In practice, we recommend for researchers to apply our Sens-Val method when the alternative is known to be less dense or when one seeks the flexibility of conducting data exploration to help design the study. At least in principle, if researchers expected to be in the dense regime, perhaps as advised by a subject matter expert, it could be beneficial to do Bonferroni correction on the full sample (assuming that they are not looking for exploration-based insights). Yet, without this prior knowledge, obtaining a well-informed notion of the sparsity of the regime would require us to split the data anyway, rendering our Sens-Val method a natural choice.}

\section{Computational Considerations}
One may contrast Method \ref{bootstrap} with the implementations of \cite{heller2009split}. In the vanilla version, they choose one outcome to test in the planning sample according to the smallest $p$-value, which is quite limited by its applicability. They also introduce a technique in which they consider all possible subsets of the outcomes and select that which has the min-max $p$-value. However, this approach is computationally infeasible for even a moderate number of outcomes, requiring the consideration of $2^L$ subsets of outcomes in the planning sample. On the other hand, Method \ref{bootstrap} is linear in the number of outcomes and is thus much more usable in practice. Moreover, due to potential correlations among the outcomes of interest, the min-max $p$-value may have complicated nuances attached to it, the theoretical properties of which are poorly understood. However, our proposed Method \ref{bootstrap} has simple theoretical properties (\ref{ourcor1}) and is robust to potential correlational complications since it relies only on the marginal structure of the outcomes. %

We also note that the bootstrap samples need not be recalculated for each $\alpha^n$ when employing the dynamic $\alpha$ framework. As the bootstrap samples are used only to obtain an estimate of $\sigma_F^l$, which is devoid of $\alpha$ from its definition via Equation \ref{alt-clt}, $\hat\sigma_F^l(\alpha)$ and $\hat\sigma_F^l(\alpha^*)$ are likely to have an $o_p(1)$ difference, leading to an $o_p({1}/{\sqrt I_{\p}})$ difference in the odds of the $l^{\text{th}}$ outcome being selected. Hence, one can safely do away with implementing the bootstrap repetitively without any additional asymptotic cost.

\pagebreak

\section{Details of Data Application} \label{dat-details}

\begin{figure}[!ht]%
\centering
\includegraphics[scale=0.5]{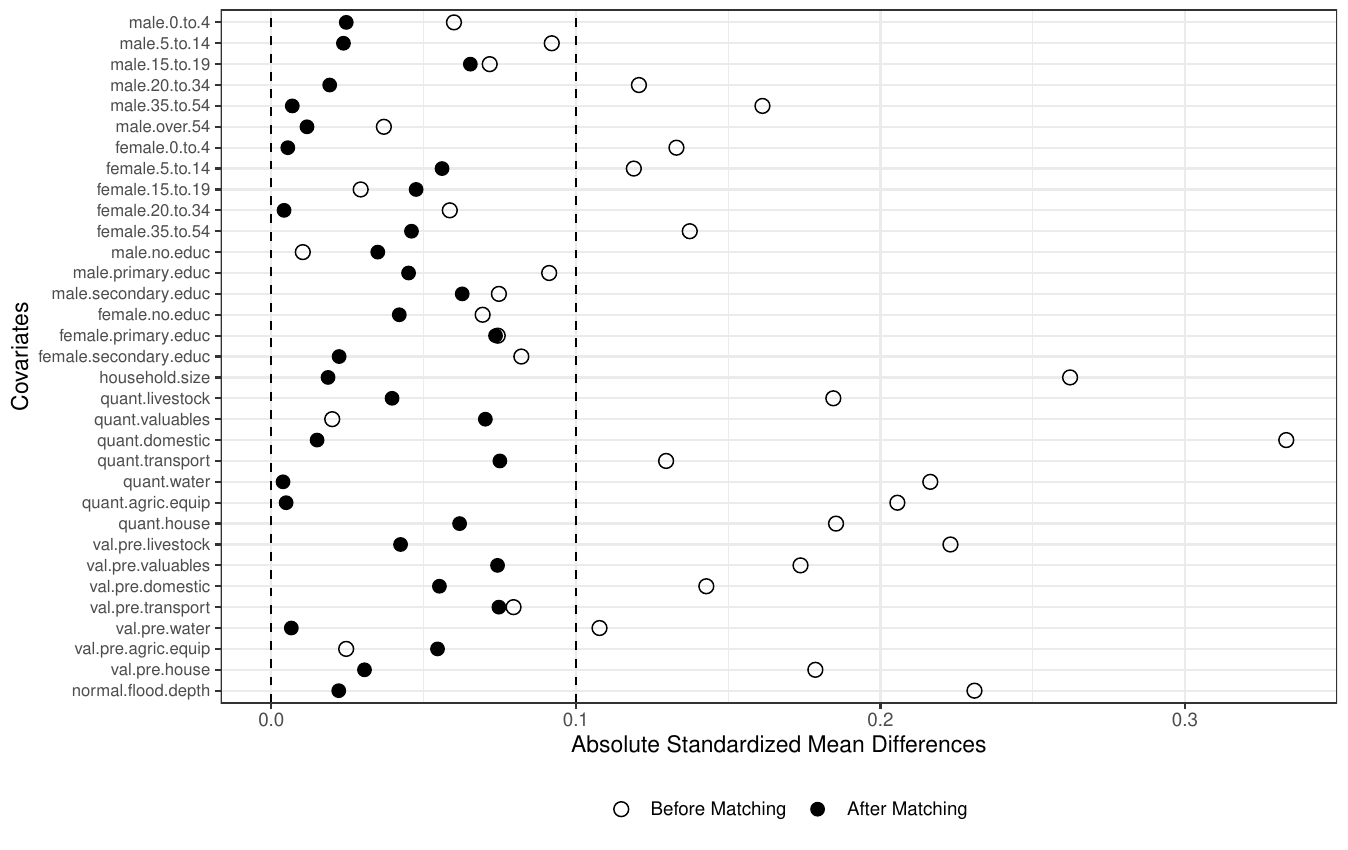}
\caption{A Love plot displaying the absolute standardized mean differences across covariates.}
\label{fig:SMDplot}
\end{figure}

\begin{table}[!ht]
\centering
\footnotesize %
\caption{Description of rejected outcomes, in accordance with the results in Table \ref{tab:results}. Among the $93$ total outcomes, only $14$ are rejected by any method at the evaluated levels of $\Gamma_{\c}.$}
\label{tab:results-legend}
\begin{tabular}{ll}
\toprule
{Outcome} & {Definition} \\
\midrule
ft.water & Feet of water in household due to flood \\
repair.cost & Cost of repairing flood damage \\
days.water & Number of days with water in household \\
days.away.home & Number of days out of home due to flood \\
tubewell.cooking & Household obtains cooking water from tubewell \\
tubewell.washing & Household obtains washing water from tubewell \\
food.price.idx & Food price index \\
food.credit.val.pc & Per capita value of credit purchases for food \\
credit.val & Total household value of credit purchases \\
days.ill & Average number of days with illness \\
pulses.budget & Share of total household expenditure used to purchase pulses \\
rice.consumed.pc & Amount of rice consumed per capita \\
food.produced.val & Total household value of food produced \\
food.produced.val.pc & Per capita value of food produced \\
\bottomrule
\end{tabular}
\end{table}

\begin{table}[!ht]
\centering
\small
\caption{All of the $93$ outcomes examined in the data application. Outcomes are grouped into flood-specific, food consumption and security, health, nutrition, and economic categories. Although many of the outcomes can be grouped into more than one category, we list them under a single group for simplicity.}
\label{tab:detailed-outcomes}
\begin{tabular}{>{\centering\arraybackslash}m{4cm} >{\raggedright\arraybackslash}m{9cm}}
\toprule
Category & Specific Outcomes \\
\midrule
Flood-Specific 
& 
days.away.home, days.water, ft.water\\
\addlinespace
Food Consumption \& Security 
& 
amt.fish.female, amt.fish.male, amt.milk.meat.egg.female, amt.milk.meat.egg.male, calories.pc, calories.per.adult, cereals.consumed.pc, cereals.budget, egg.consumed.pc, egg.budget, fish.consumed.pc, fish.budget, food.price.idx, fruit.consumed.pc, fruit.budget, icrs.fish.female, icrs.fish.male, meat.consumed.pc, meat.budget, milk.consumed.pc, milk.budget, oil.consumed.pc, oil.budget, protein.pc, protein.per.adult, pulses.consumed.pc, pulses.budget, rice.consumed.pc, rice.budget, veg.consumed.pc, veg.budget, wheat.consumed.pc, wheat.budget \\
\addlinespace
Health & 
any.ill.0-5yrs, any.ill.6-18yrs, any.ill.fever, any.ill.female, any.ill.gastrointest, any.ill.male, any.ill.over18yrs, any.ill.respiratory, days.ill, distance.cooking, distance.drinking, distance.washing, liters.water.collection, sanitary.latrine, time.water.collection, tubewell.cooking, tubewell.drinking, tubewell.washing \\
\addlinespace
Nutrition & 
rd3.bmi, rd3.bmi.adolesc.women, rd3.height.on.age, rd3.height.on.age.preschool, rd3.weight.on.age, rd3.weight.on.age.preschool, rd3.weight.on.height, rd3.weight.on.height.preschool \\
\bottomrule
\end{tabular}
\end{table}

\newpage

\subsection{Summary Statistics on Planning Sample}

\small
\begin{longtable}{llrr}
\caption{Summary Statistics for Numeric Variables} \\
\hline
&
Statistic&
Control&
Treated
\\\hline
\textbf{adult.equivalent.tot.expend} & & & \\\hline

 & Min. & 348.49 & 340.62 \\
 & 1st Qu. & 605.52 & 659.45 \\
 & Median & 768.45 & 739.21 \\
 & Mean & 860.16 & 1005.66 \\
 & 3rd Qu. & 1056.41 & 1137.36 \\
\hline
\end{longtable}

\begin{longtable}{llrr}
\caption{Proportions for Binary Variables} \\
\hline
&
Value&
Control&
Treated
\\\hline
\textbf{any.electricity} & & & \\\hline

 & 0 & 0.82 & 0.90 \\
 & 1 & 0.18 & 0.10 \\
\textbf{any.ill.0-5yrs} & & & \\\hline

 & 0 & 0.77 & 0.65 \\
 & 1 & 0.23 & 0.35 \\
\textbf{any.ill.6-18yrs} & & & \\\hline

 & 0 & 0.77 & 0.74 \\
 & 1 & 0.23 & 0.26 \\
\textbf{any.ill.female} & & & \\\hline

 & 0 & 0.44 & 0.35 \\
 & 1 & 0.56 & 0.65 \\
\textbf{any.ill.fever} & & & \\\hline

 & 0 & 0.65 & 0.53 \\
 & 1 & 0.35 & 0.47 \\
\textbf{any.ill.gastrointest} & & & \\\hline

 & 0 & 0.50 & 0.59 \\
 & 1 & 0.50 & 0.41 \\
\textbf{any.ill.male} & & & \\\hline

 & 0 & 0.40 & 0.42 \\
 & 1 & 0.60 & 0.58 \\
\textbf{any.ill.over18yrs} & & & \\\hline

 & 0 & 0.30 & 0.30 \\
 & 1 & 0.70 & 0.70 \\
\textbf{any.ill.respiratory} & & & \\\hline

 & 0 & 0.97 & 0.79 \\
 & 1 & 0.03 & 0.21 \\
\textbf{sanitary.latrine} & & & \\\hline

 & 0 & 0.81 & 0.88 \\
 & 1 & 0.19 & 0.12 \\
\textbf{tubewell.cooking} & & & \\\hline

 & 0 & 0.16 & 0.49 \\
 & 1 & 0.84 & 0.51 \\
\textbf{tubewell.drinking} & & & \\\hline

 & 0 & 0.02 & 0.07 \\
 & 1 & 0.98 & 0.93 \\
\textbf{tubewell.washing} & & & \\\hline

 & 0 & 0.37 & 0.63 \\
 & 1 & 0.63 & 0.37 \\\hline
\end{longtable}

\newpage

\subsection{Summary Statistics on Analysis Sample}

\begin{longtable}{llrr}
\caption{Summary Statistics for Numeric Variables} \\
\hline
&
Statistic&
Control&
Treated
\\\hline
\textbf{adult.equivalent.tot.expend} & & & \\\hline

 & Min. & 171.06 & 158.64 \\
 & 1st Qu. & 682.88 & 686.60 \\
 & Median & 922.55 & 912.32 \\
 & Mean & 1042.89 & 1130.61 \\
 & 3rd Qu. & 1236.91 & 1379.69 \\
\hline
\end{longtable}

\newpage

\begin{longtable}{llrr}
\caption{Proportions for Binary Variables} \\
\hline
&
Value&
Control&
Treated
\\\hline
\textbf{any.electricity} & & & \\\hline

 & 0 & 0.85 & 0.83 \\
 & 1 & 0.15 & 0.17 \\
\textbf{any.ill.0-5yrs} & & & \\\hline

 & 0 & 0.73 & 0.72 \\
 & 1 & 0.27 & 0.28 \\
\textbf{any.ill.6-18yrs} & & & \\\hline

 & 0 & 0.81 & 0.68 \\
 & 1 & 0.19 & 0.32 \\
\textbf{any.ill.female} & & & \\\hline

 & 0 & 0.53 & 0.44 \\
 & 1 & 0.47 & 0.56 \\
\textbf{any.ill.fever} & & & \\\hline

 & 0 & 0.59 & 0.54 \\
 & 1 & 0.41 & 0.46 \\
\textbf{any.ill.gastrointest} & & & \\\hline

 & 0 & 0.62 & 0.59 \\
 & 1 & 0.38 & 0.41 \\
\textbf{any.ill.male} & & & \\\hline

 & 0 & 0.53 & 0.37 \\
 & 1 & 0.47 & 0.63 \\
\textbf{any.ill.over18yrs} & & & \\\hline

 & 0 & 0.51 & 0.38 \\
 & 1 & 0.49 & 0.62 \\
\textbf{any.ill.respiratory} & & & \\\hline

 & 0 & 0.91 & 0.89 \\
 & 1 & 0.09 & 0.11 \\
\textbf{sanitary.latrine} & & & \\\hline

 & 0 & 0.85 & 0.87 \\
 & 1 & 0.15 & 0.13 \\
\textbf{tubewell.cooking} & & & \\\hline

 & 0 & 0.18 & 0.51 \\
 & 1 & 0.82 & 0.49 \\
\textbf{tubewell.drinking} & & & \\\hline

 & 0 & 0.02 & 0.05 \\
 & 1 & 0.98 & 0.95 \\
\textbf{tubewell.washing} & & & \\\hline

 & 0 & 0.38 & 0.63 \\
 & 1 & 0.62 & 0.37 \\\hline
\end{longtable}

\newpage

\subsection{Summary Statistics on Entire Sample}

\begin{longtable}{llr}
\caption{Summary Statistics for Numeric Variables} \\
\hline
&
Statistic&
Sample
\\\hline
\textbf{adult.equivalent.tot.expend} & & \\\hline

 & Min. & 158.64 \\
 & 1st Qu. & 667.60 \\
 & Median & 898.24 \\
 & Mean & 1055.98 \\
 & 3rd Qu. & 1234.03 \\
\hline
\end{longtable}

\newpage

\begin{longtable}{llr}
\caption{Proportions for Binary Variables} \\
\hline
&
Value&
Sample
\\\hline
\textbf{any.electricity} & & \\\hline

 & 0 & 0.84 \\
 & 1 & 0.16 \\
\textbf{any.ill.0-5yrs} & & \\\hline

 & 0 & 0.72 \\
 & 1 & 0.28 \\
\textbf{any.ill.6-18yrs} & & \\\hline

 & 0 & 0.75 \\
 & 1 & 0.25 \\
\textbf{any.ill.female} & & \\\hline

 & 0 & 0.47 \\
 & 1 & 0.53 \\
\textbf{any.ill.fever} & & \\\hline

 & 0 & 0.57 \\
 & 1 & 0.43 \\
\textbf{any.ill.gastrointest} & & \\\hline

 & 0 & 0.59 \\
 & 1 & 0.41 \\
\textbf{any.ill.male} & & \\\hline

 & 0 & 0.44 \\
 & 1 & 0.56 \\
\textbf{any.ill.over18yrs} & & \\\hline

 & 0 & 0.42 \\
 & 1 & 0.58 \\
\textbf{any.ill.respiratory} & & \\\hline

 & 0 & 0.90 \\
 & 1 & 0.10 \\
\textbf{sanitary.latrine} & & \\\hline

 & 0 & 0.86 \\
 & 1 & 0.14 \\
\textbf{tubewell.cooking} & & \\\hline

 & 0 & 0.34 \\
 & 1 & 0.66 \\
\textbf{tubewell.drinking} & & \\\hline

 & 0 & 0.04 \\
 & 1 & 0.96 \\
\textbf{tubewell.washing} & & \\\hline

 & 0 & 0.50 \\
 & 1 & 0.50 \\\hline
\end{longtable}

\normalsize

\end{document}